\documentclass[english,showpacs,twocolumn, superscriptaddress]{revtex4-1}
\usepackage[T1]{fontenc}
\usepackage[latin9]{inputenc}
\usepackage{amsmath}
\usepackage{graphicx}
\usepackage{wasysym}

\makeatletter
\usepackage[unicode=true,pdfusetitle,
 bookmarks=true,bookmarksnumbered=false,bookmarksopen=false,
 breaklinks=false,pdfborder={0 0 0},pdfborderstyle={},backref=false,colorlinks=true]
 {hyperref}
\hypersetup{
 urlcolor=blue, citecolor=blue}

\makeatother

\usepackage{babel}
\begin{document}
\title{Effects of Postselected von Neumann Measurement on the Properties
of Single-Mode Radiation Fields}
\author{Yusuf Turek}
\date{\today}
\email{yusufu1984@hotmail.com}

\address{School of Physics and Electronic Engineering, Xinjiang Normal University,
Urumqi, Xinjiang 830054, China}
\begin{abstract}
Postselected von Neumann measurement characterized by postselection
and weak value has been found potential applications in quantum metrology
and solved plenty of fundamental problems in quantum theory. As an
application of this new measurement technique in quantum optics and
quantum information processing, its effects on the features of single-mode
radiation fields such as coherent state, squeezed vacuum state and
Schrödinger cat sate are investigated by considering full-order effects
of unitary evolution. The results show that the conditional probabilities
of finding photons, second-order correlation functions, $Q_{m}$-factors
and squeezing effects of those states after postselected measurement
significantly changed compared with the corresponding initial pointer
states. 
\end{abstract}
\maketitle

\section{Introduction}

Most researches in von Neumann type quantum measurement in recent
years has focused on postseleceted weak measurement with sufficiently
weak coupling between the measured system and pointer since it is
useful to study the nature of the quantum world. This kind of postsselected
weak measurement theory is originally proposed by Aharonov, Albert,
and Vaidman in 1988 \citep{Aharonov1988}, and considered as a generalized
version of standard von-Neumann type measurement \citep{Edwards1955The}.
The result of weak coupling postselected weak measurement is called
``weak value'', and generally is a complex number. One of the special
feature of the ``weak value '' is that it can take the values, which
lie beyond the normal eigenvalue range of corresponding observable,
and this effect is very clear if the pre- and post-selected state
almost orthogonal. This feature of ``weak value '' is called signal
amplification property of postselected weak measurement and its first
weak signal amplification property experimentally demonstrated in
1991 \citep{Ritchie1991}. After that, it has been widely used and
elucidated tremendous fundamental problems in quantum mechanics. For
details about the weak measurement and its applications in signal
amplification processes, we refer the reader to the recent overview
of the field \citep{KOFMAN201243,RevModPhys.86.307}. As we mentioned
earlier, in postselected weak measurement the interaction strength
is weak, and it is enough to consider up to the first order evolution
of unitary operator for the whole measurement processes. However,
if we want to connect the weak and strong measurement, check to clear
the measurement feedback of postselected weak measurement and analyze
experimental results obtained in nonideal measurements, the full order
evolution of unitary operator is needed \citep{PhysRevA.72.052111,PhysRevA.77.042108,PhysRevA.85.022122},
we call this kind of measurement is postselected von-Neumann measurement.
We know that in some quantum metrology problems the precision of the
measurement depends on measuring devices and require to optimize the
pointer states. The merits of postselected von-Neumann measurement
can be seen in pointer optimization schemes. 

Recently, the state optimization problem in postselected von-Neumann
measurement have been presented widely, such as taking the Gaussian
states \citep{KOFMAN201243,PhysRevA.85.012113}, Hermite-Gaussian
or Laguerre-Gaussian states, and non-classical states \citep{Turek_2015,DELIMABERNARDO2014194}.
The advantages of non-classical pointer states in increasing postselected
measurement precision have been examined in recent studies \citep{PhysRevA.92.022109,Turek_2015,PhysRevLett.115.120401}.
Furthermore, in Ref. \citep{Turek2018}, the authors studied the effects
of postselected measurement characterized by modular value \citep{Vaidman2010}
to show the properties of semi-classical and non-classical pointer
states considering the coherent, coherent squeezed, and Schrodinger
cat state as a pointer. Since in their scheme the pointer operator
is a projection operator onto one of the states of the basis of the
pointer's Hilbert space and interaction strength only has taken one
definite value, it can't completely describe the effects of postselected
measurement on the properties of the radiation field. However, to
the knowledge of the author, the issue related to the study of the
effects of postselected von-Neumann measurement considering all interaction
strengths to the inherent properties such as photon distribution,
photon statistics, and squeezing effects of radiation field have not
been handled and need to be investigated. 

Furthermore, the above mentioned properties of quantum radiation fields
have many useful technological applications in quantum optics and
quantum information processing such as single photon generation and
detection \citep{Buller2010Single}, gravitational wave detection
\citep{PhysRevD.80.042006,Grote2016High}, quantum teleportation,
\citep{PhysRevLett.80.869,PhysRevA.60.937,PhysRevA.64.052308,PhysRevA.64.022313},
quantum computation \citep{PhysRevA.68.042319}, generation and manipulation
of atom-light entanglement \citep{PhysRevA.73.062329,LiEntanglement,Hacker},
and precision measurements \citep{PhysRevA.66.023819} etc. We know
that the realization of these processes depends on the optimization
of the related input quantum states such as coherent state \citep{PhysRev.131.2766},
squeezed state \citep{Carranza:12,Andersen_2016} and Schrödinger
cat state \citep{monroe1996a,OurjoumtsevGeneration,RevModPhys.85.1083}.
As introduced in previous part, the postselected weak measurement
technique have more advantageous on signal optimization problem \citep{PhysRevA.80.041803}
and precision metrology \citep{Hosten787,PhysRevLett.102.173601,PhysRevA.88.053815,PhysRevA.82.063822,PhysRevLett.112.200401,Pfeifer:11,DELIMABERNARDO20142029,Viza:13,Egan:12}
than traditional measurement theory. Thus, the investigation of the
effects of postselected von-Neumann measurement on the statistical
and squeezing properties of radiation fields is worth to study to
optimize the related quantum states to provide more effective methods
for the implementation of the above mentioned technological processes.

In this paper, motivated by the above listed problems, we describe
and examine the effects of postselected von-Neumann measurement characterized
by postselection and weak value on the properties of single-mode radiation
fields. In order to achieve our goal, we choose the three typical
states such as coherent state, squeezed vacuum state, and Schrödinger
cat state as pointers and consider their polarization degree of freedom
as measured system, respectively. By taking full-order evolution of
the unitary operator of our system, we separately study the photon
distributions, statistical properties and squeezing effects of radiation
fields and compare that with the initial state cases. To give more
details about the effects of postselected von-Neumann measurement
on radiation fields, we plot more figures by considering all parameters
which related to the properties of those radiation fields, respectively.
This study shows that the postselected measurement changes the properties
of single-mode radiation fields dramatically for strong and weak measurement
regimes with specific weak values. Especially the photon statistics
and squeezing effect of coherent pointer state is too sensitive for
the postselected von-Neumann measurement processes.

This paper is organized as follows: In Section. \ref{sec:II}, we
describe the model of our theory by brief reviewing the postselected
von-Neumann measurement and outline the tasks we want to investigate.
In Section. \ref{sec:III}, we separately calculate exact analytical
expressions of photon distribution, second-order correlation function,
Mandel factor and squeezing parameter of those three-pointers by taking
into account the full-order evaluation of unitary operator under final
states which given after postselected measurement finished and present
our main results, respectively. Finally, we summarize our findings
of this study in Sec. \ref{sec:IV}. 

\section{\label{sec:II}Model Setup}

To build up our model, we begin to review some basic concepts of the
von-Neumann measurement theory. The standard measurement consists
of three elements; measured system, pointer ( measuring device or
meter) and environment, which induce the interaction between system
and pointer. The description of the measurement process can be written
in terms of Hamiltonian, and the total Hamiltonian of a measurement
composed of three parts \citep{Aharonov2005}
\begin{equation}
H=H_{s}+H_{p(m)}+H_{int},\label{eq:eq}
\end{equation}
where $H_{s}$ and $H_{p(m)}$ represents the Hamiltonian of measured
system and measuring device, respectively, and $H_{int}$ is the interaction
Hamiltonian between measured system and pointer. By considering the
measurement efficiency and accuracy, in standard quantum measurement
theory the interaction time between system and meter required to be
too short so that the $H_{s}$ and $H_{p(m)}$ doesn't affect the
final readout of the measurement result. Thus, in general, a measurement
can only be described by the interaction part $H_{int}$ of the total
Hamiltonian, and it is taken to the standard von Neumann Hamiltonian
as \citep{Edwards1955The}

\begin{equation}
H_{int}=g(t)\hat{A}\hat{P},\ \ \ \ \int_{t_{0}}^{t}g(t)dt=g\text{\ensuremath{\delta(t-t_{0})}.}\label{eq:Hamil}
\end{equation}
Here, $\text{\ensuremath{\hat{A}}}$ represents the Hermitian operator
corresponding to the observable of the system that we want to measure
with $\hat{A}\vert\phi_{n}\rangle=a_{n}\vert\phi_{n}\rangle$, and
$\hat{P}$ is the conjugate momentum operator to the position operator
$\hat{X}$ of the pointer, i.e., $[\hat{X},\hat{P}]=i\hat{I}$. The
coupling $g(t)$ is a nonzero function in a finite interaction time
interval $t-t_{0}$. 

One can express the position $\hat{X}$ and momentum operator $\hat{P}$
of the pointer by using the annihilation $\hat{a}$ and creation operator
$\hat{a}^{\dagger}$ (satisfying $[\hat{a},\hat{a}^{\dagger}]=1$)
as \citep{GerryIntroductory} 
\begin{eqnarray}
\hat{X} & = & \sigma(\hat{a}^{\dagger}+\hat{a}),\label{eq:annix}\\
\hat{P} & = & \frac{i}{2\sigma}(\hat{a}^{\dagger}-\hat{a}),\label{eq:anniy}
\end{eqnarray}
where $\sigma$ is the width of the beam. The Hamiltonian $H_{int}$
can be rewritten in terms of $a$ and $a^{\dagger}$ as $H_{int}=\frac{ig(t)}{2\sigma}\hat{A}(a^{\dagger}-a)$.
If the Hermitian operator $\hat{A}$ satisfies the property $\hat{A}^{2}=\hat{I}$,
then the unitary evolution operator $e^{-i\int_{t_{0}}^{t}H_{int}d\tau}$
of our coupling system is given by 
\begin{equation}
e^{-ig\hat{A}\otimes\hat{P}}=\frac{1}{2}(\hat{I}+\hat{A})\otimes D(\frac{s}{2})+\frac{1}{2}(\hat{I}-\hat{A})\otimes D(-\frac{s}{2}),\label{eq:UNA1}
\end{equation}
where $s$ is defined by $s:\equiv g/\sigma$, and $D\left(\frac{s}{2}\right)$
is a displacement operator and its expression can be written as $D(\frac{s}{2})=e^{\frac{s}{2}(\hat{a}^{\dagger}-\hat{a})}$.
We have to note that $s$ characterizes the measurement strength,
and $s\ll1$$(s>1)$ corresponds to weak (strong) measurement regimes,
respectively. 

Assume that the system initially prepared in the state $\vert\psi_{i}\rangle$
and the initial pointer state is $\vert\phi\rangle$, then after interaction
we project the state $e^{-i\int_{t_{0}}^{t}H_{int}d\tau}\vert\psi_{i}\rangle\vert\phi\rangle$
onto the postselected system state $\vert\psi_{f}\rangle$, we can
obtain the final state of the pointer. Furthermore, the normalized
final state of the pointer can be written as 

\begin{equation}
\vert\Phi\rangle=\mathcal{N\!}\left[\!\frac{1}{2}(\hat{I}\!+\!\langle A\rangle{}_{w}\!)\!\otimes\!D(\frac{s}{2})\!+\!\frac{1}{2}(\hat{I}\!-\!\langle A\rangle{}_{w})\!\otimes\!D(-\frac{s}{2})\!\right]\!\!\vert\phi\rangle.\label{eq:6}
\end{equation}
Here, $\mathcal{N}$ is the normalization coefficient of $\vert\Phi\rangle$,
and 
\begin{equation}
\langle A\rangle{}_{w}=\frac{\langle\psi_{f}\vert\hat{A}\vert\psi_{i}\rangle}{\langle\psi_{f}\vert\psi_{i}\rangle},\label{eq:WV}
\end{equation}
is called the weak value of Hermitian operator $\hat{A}$. From Eq.
(\ref{eq:WV}), we know that when the preselected state $\vert\psi_{i}\rangle$
and the postselected state $\vert\psi_{f}\rangle$ of the system are
almost orthogonal, the absolute value of the weak value can be arbitrarily
large and can beyond the eigenvalue region of observable $A$. This
feature leads to weak value amplification of weak signals, and solved
a lot of important problems in physics.

In this study, we take the transverse spatial degree of freedom of
single-mode radiation field as pointer and its polarization degree
of freedom as a measured system. We suppose that the operator to be
observed is the spin $x$ component of a spin- $1/2$ particle, i.e.,
\begin{equation}
A=\sigma_{x}=\vert\uparrow_{z}\rangle\langle\downarrow_{z}\vert+\vert\downarrow_{z}\rangle\langle\uparrow_{z}\vert,
\end{equation}
where $\vert\uparrow_{z}\rangle$ and $\langle\downarrow_{z}\vert$
are eigenstates of the $z$-component of spin, $\sigma_{z}$, with
the corresponding eigenvalues $1$ and $-1$, respectively. We assume
that the pre- and post-selected states of measured system are 
\begin{equation}
\vert\psi_{i}\rangle=\cos\frac{\theta}{2}\vert\uparrow_{z}\rangle+e^{i\varphi}\sin\frac{\theta}{2}\vert\downarrow_{z}\rangle,\label{eq:Pre}
\end{equation}
and 
\begin{equation}
\vert\psi_{f}\rangle=\vert\uparrow_{z}\rangle,\label{eq:Post}
\end{equation}
respectively, and the corresponding weak value, Eq. (\ref{eq:WV})
reads 
\begin{equation}
\langle\sigma_{x}\rangle_{w}=e^{i\varphi}\tan\frac{\theta}{2},\label{eq:WV-1}
\end{equation}
where $\theta\in[0,\pi]$ and $\varphi\in[0,2\pi)$. Here, in our
system, the post-selection probability is $P_{s}=\cos^{2}\frac{\theta}{2}$.
Note that throughout the rest of this study we will use this weak
value for our purposes.

Furthermore, we take the initial pointer state $\vert\phi\rangle$
as a coherent state, squeezed vacuum state and Schrodinger cat state
to study the effects of postselected measurement on the properties
of those pointers, respectively. To achieve our goal:

1. we study the conditional probability of finding $n$ photons after
postselected measurement. For the state $\vert\Phi\rangle$, the conditional
probability of finding $n$ photons can be calculated by 
\begin{equation}
P_{post}(n)=\vert\langle n\vert\Phi\rangle\vert^{2},\label{eq:12}
\end{equation}
and we compare it with the probability $P(n)=\vert\langle n\vert\phi\rangle\vert^{2}$
of initial pointer state $\vert\phi\rangle$.

2.we investigate the second-order correlation function $g^{(2)}(0)$
and Mandel factor $Q_{m}$ for the $\vert\Phi\rangle$ state. The
second-order correlation function of a single-mode radiation field
is defined as 
\begin{align}
g^{(2)}(0) & =\frac{\langle a^{\dagger}a^{\dagger}aa\rangle}{\langle a^{\dagger}a\rangle^{2}},\label{eq:13}
\end{align}
and the Mandel factor $Q_{m}$of the radiation field can be expressed
in terms of $g^{(2)}(0)$ as 
\begin{equation}
Q_{m}=\langle n\rangle\left[g^{(2)}(0)-1\right]\label{eq:14}
\end{equation}
 If $0\le g^{(2)}(0)<1$ and $-1\le Q_{m}<0$ simultaneously, the
corresponding radiation field have sub-Poissonian statistics and more
nonclassical. We have mention that the quantity $Q_{m}$ can never
be smaller than -1 for any radiation field, and negative $Q_{m}$
values which equivalent to sub-poissonian statistics, cannot be produced
by any classical field \citep{Mandel1979Sub}.

3.we check the squeezing parameter of single-mode radiation field
for the $\vert\Phi\rangle$ state. The squeezing parameter of the
single-mode radiation field reads as \citep{Agarwal2013}

\begin{equation}
S_{\phi}=\langle:X_{\phi}^{2}:\rangle-\langle X_{\phi}\rangle^{2}\label{eq:15}
\end{equation}
where 
\begin{equation}
X_{\phi}=\frac{1}{\sqrt{2}}\left(ae^{-i\phi}+a^{\dagger}e^{i\phi}\right),\ \ \ [X_{\phi},X_{\phi+\frac{\pi}{2}}]=i\label{eq:16}
\end{equation}
is quadrature operator of the field, and $:\ :$ stands for the normal
ordering of the operator defined by $:aa^{\dagger}:=a^{\dagger}a$,
whereas $aa^{\dagger}=a^{\dagger}a+1$. We can see that $S_{\theta}$
is related to the variance of $X_{\theta},$i.e. 
\begin{equation}
S_{\phi}=\left(\triangle X_{\phi}\right)^{2}-\frac{1}{2}\label{eq:17}
\end{equation}
where $\triangle X_{\phi}=\sqrt{\langle X_{\phi}^{2}\rangle-\langle X_{\phi}\rangle^{2}}$
. The minimum value of $S_{\phi}$ is $-\frac{1}{2}$ and for a nonclassical
state $S_{\phi}\in[-\frac{1}{2},0)$. 

In the next section, we will study the above properties of three typical
radiation fields by taking into the postselected measurement with
arbitrary interaction strengths and weak values, respectively.

\section{\label{sec:III}Effects of postselected measurement to typical single-mode
radiation fields}

In this section, we study the statistical properties and squeezing
effects of typical single-mode radiation fields such as a coherent
state, squeezed vacuum state and Schrödinger cat state, respectively,
and compare the results with the corresponding initial state's case. 

\subsection{coherent state}

The coherent state is typical semi-classical, quadrature minimum-uncertainty
state for all mean photon numbers. The coherent state was originally
introduced by Schrödinger in 1926 as a Gaussian beam to describe the
evolution of a harmonic oscillator \citep{Schr1926Der}, and the mathematical
formulation of coherent state (also called Glauber state) has been
introduced by Roy J. Glauber in 1963 \citep{PhysRev.131.2766}. Coherent
state play an important role in representing quantum dynamics, especially
when the quantum evolution is close to classical \citep{Klauder1985Coherent,Perelomov1986Generalized}
. Here, we take the coherent state as initial pointer state \citep{Scullyl1997}
\begin{equation}
\left|\alpha\right\rangle =D(\alpha)\left|0\right\rangle ,\label{eq:Coh sta}
\end{equation}
where $D(\alpha)=e^{\alpha a^{\dagger}-\alpha^{\ast}a}$, and $\alpha=re^{i\vartheta}$
is an arbitrary complex number. After unitary evolution that is given
in Eq. (\ref{eq:UNA1}), the total system state is post-selected to
$\vert\psi_{f}\rangle$, then we obtain the following normalized final
state of coherent pointer state:
\begin{eqnarray}
\vert\Psi\rangle & = & \frac{\lambda}{\sqrt{2}}\times\label{eq:GeC}\\
 &  & [(1+\!\!\langle A\rangle_{w})e^{-i\frac{s}{2}Im(\alpha)}\vert\alpha\!\!+\frac{s}{2}\rangle+(1-\!\!\langle A\rangle_{w})e^{i\frac{s}{2}Im(\alpha)}\vert\alpha\!\!-\frac{s}{2}\rangle],\nonumber 
\end{eqnarray}
where the normalization coefficient is given as 
\begin{eqnarray}
\lambda^{-2} & = & 1+\vert\langle A\rangle_{w}\vert^{2}\nonumber \\
\!\! & \!\! & \!\!+e^{-\frac{1}{2}s^{2}}\!\!Re[(1\!\!+\langle A\rangle_{w})(1\!\!-\langle A\rangle_{w}^{\ast})e^{-2isIm(\alpha)}],\label{eq:Co1}
\end{eqnarray}
and $Im$ ($Re$) represents the imaginary (real) part of a complex
number.

\begin{figure}
\includegraphics[width=4cm]{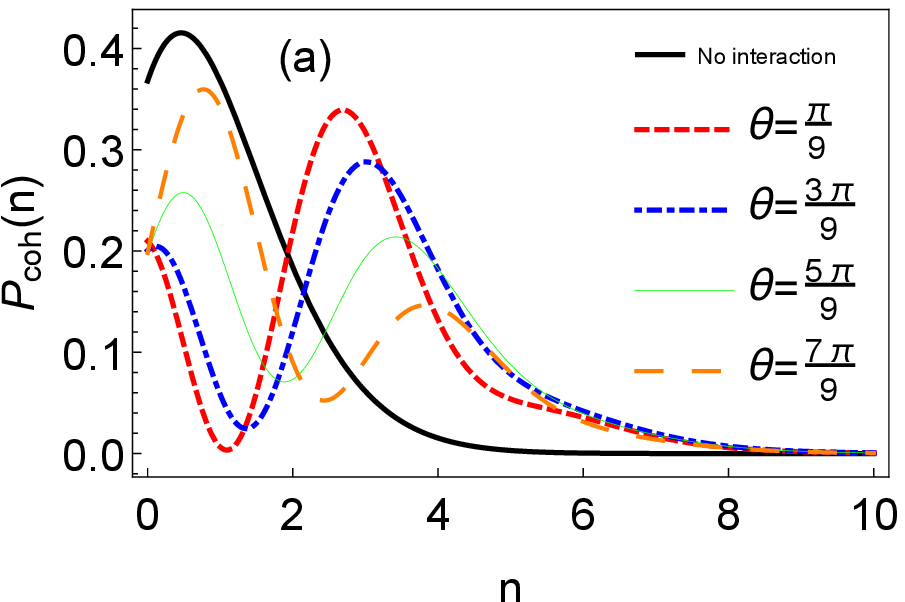}\includegraphics[width=4cm]{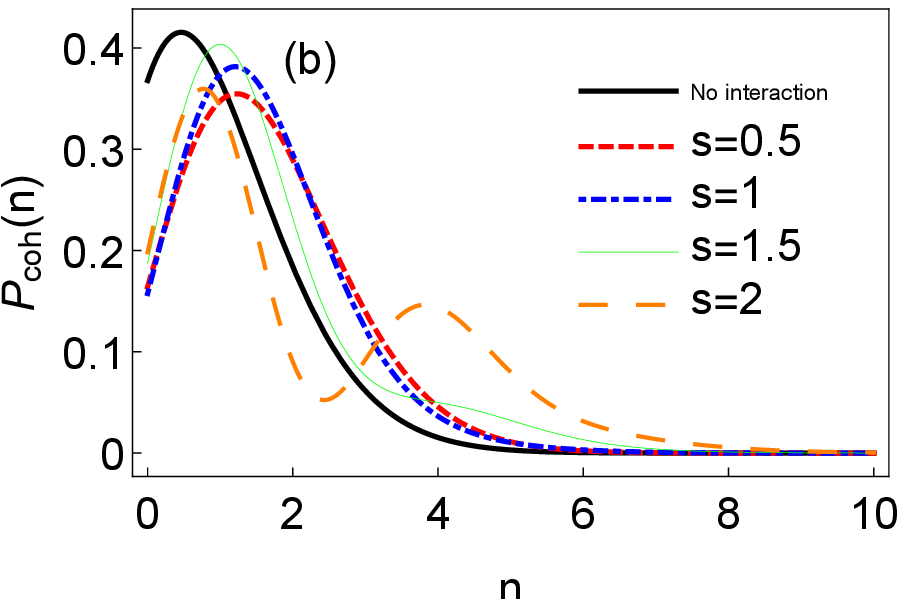}

\caption{\label{fig:fig1}(Color online) Photon distribution $P_{coh}(n)$
of coherent state as a function of photon number $n$. $\vartheta=\frac{\pi}{3}$,
$\varphi=\frac{\pi}{4}$, $r=1$. (a) $P_{coh}(n)$ plotted for $s=2,$
and for various weak value ($\langle\sigma_{x}\rangle_{w}=e^{\frac{i\pi}{3}}\tan\frac{\theta}{2}$)
and no interaction case (black curve). (b) $P_{coh}(n)$ plotted for
$\theta=\frac{7\pi}{9}$, and for various coupling strength $s$.}
\end{figure}

The conditional probability of finding $n$ photons after taking a
postselected measurement for state $\vert\Psi\rangle$ is obtained
from Eq. (\ref{eq:12}) by changing $\vert\Phi\rangle$ to a specific
normalized state $\vert\Psi\rangle$. We plot the conditional probability
of finding $n$ photons under the normalized state $\vert\Psi\rangle$
and the analytical results are shown in Fig. \ref{fig:fig1}. As indicated
in Fig. \ref{fig:fig1}, the black solid line represents the photon
distribution probability for the initial state of coherent pointer
state before postselected measurement, and it has a Poisson probability
distribution. However, as shown in Fig. \ref{fig:fig1} the postselected
measurement can change the photons Poisson distribution with increasing
the weak value and changing the interaction strength $s$. Furthermore,
from Fig. \ref{fig:fig1}(a) we can know that $P_{coh}(n)$ is larger,
in postslected case for some small photon numbers regions rather than
no interaction case. 

\begin{figure}
\includegraphics[width=4cm]{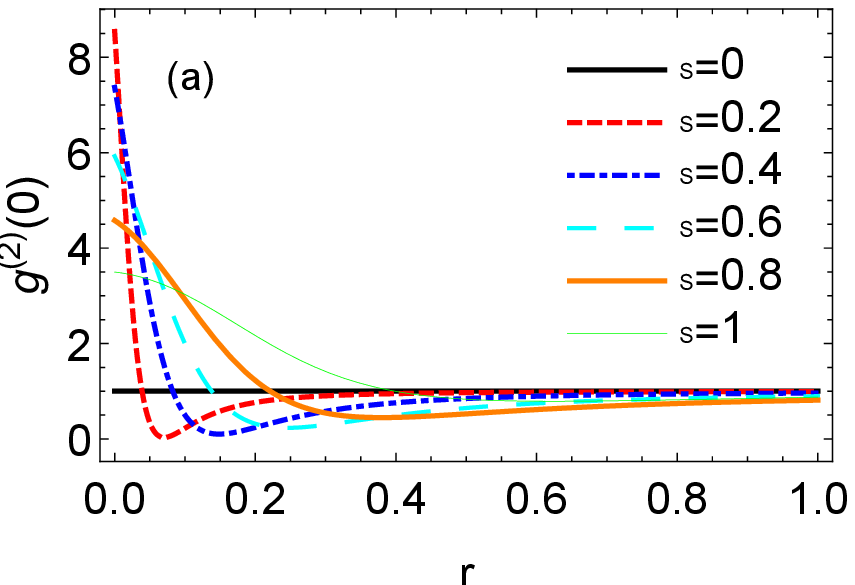}\includegraphics[width=4cm]{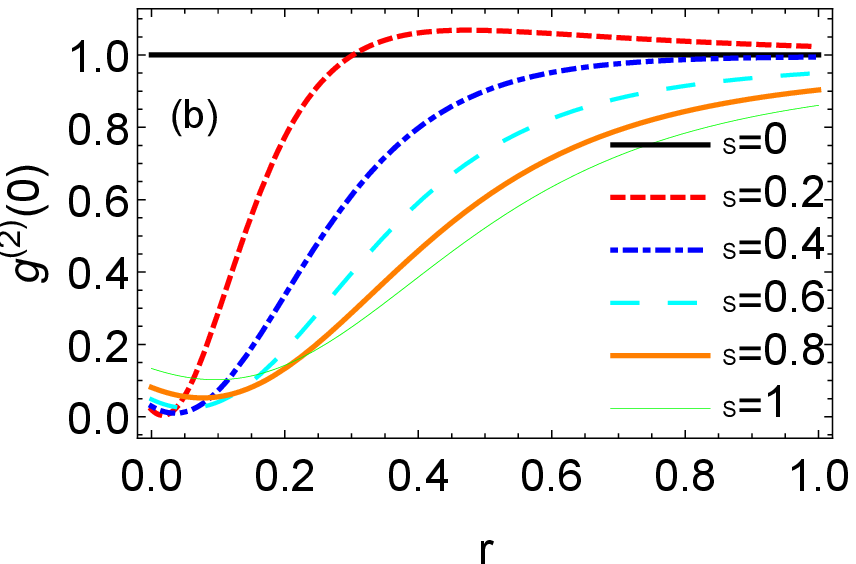}

\includegraphics[width=4cm]{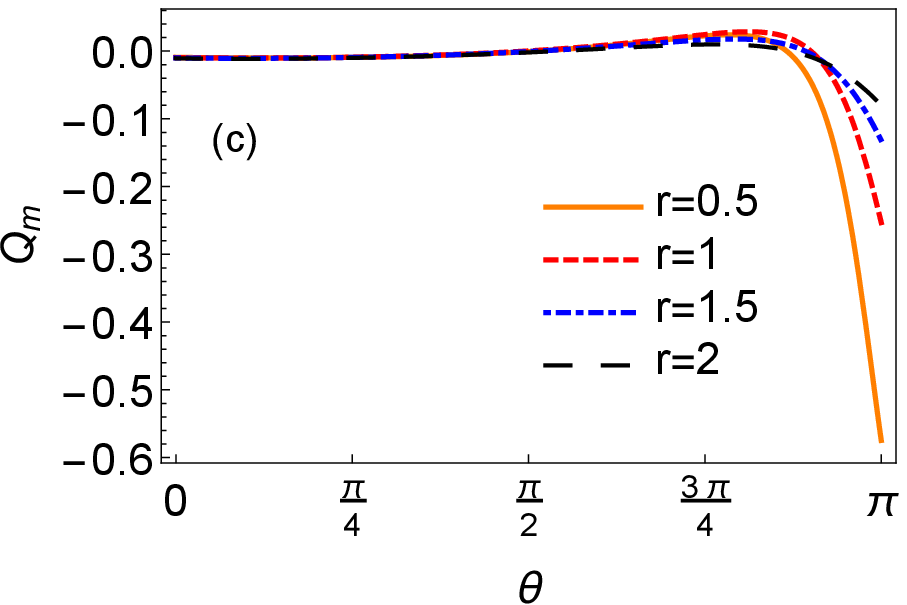}\includegraphics[width=4cm]{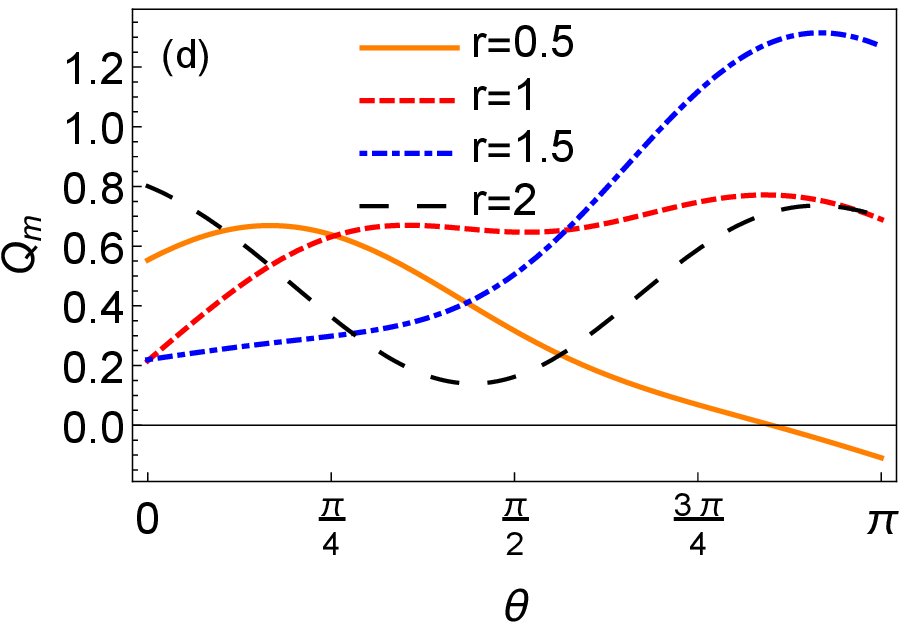}

\caption{\label{fig:2}(Color online) Second-order correlation function $g_{coh}^{(2)}(0)$
and Mandel factor $Q_{m,coh}$ for coherent state after postselected
measurement. Here, $\varphi=\frac{4\pi}{5}$, $\vartheta=\frac{\pi}{3}$.
$g_{coh}^{(2)}(0)$ vs coherent state parameter $r$ for different
interaction strength $s$ (black sold line represents for initial
coherent state $\vert\alpha\rangle$ case) and for (a) $\text{\ensuremath{\theta=\frac{\pi}{3}}}$,
or (b) $\theta=\frac{7\pi}{9}$ . Mandel factor $Q_{m,coh}$ plotted
as a function of weak value for different $r$ of coherent state and
for (c) $s=0.2$, or (d) $s=2$.}
\end{figure}

For coherent state $\vert\alpha\rangle$ (Eq. (\ref{eq:Coh sta}))
the second-order correlation function $g^{(2)}(0)$ is equal to one,
i.e.,$g^{(2)}(0)=1$ and the Mandel factor is equal to zero. Now we
will calculate the $g_{coh}^{(2)}(0)$ and $Q_{m,coh}$ considering
the final pointer state $\vert\Psi\rangle$ for coherent pointer state.The
expectation value of photon number operator $\hat{n}=a^{\dagger}a$
under the state $\vert\Psi\rangle$ is 
\begin{align}
\langle n\rangle_{\Psi} & =\frac{\vert\lambda\vert^{2}}{4}\{\text{\ensuremath{\vert1+\langle A\rangle_{w}\vert^{2}\vert\alpha+\frac{s}{2}\vert^{2}+\vert1-\langle A\rangle_{w}\vert^{2}\vert\alpha-\frac{s}{2}\vert^{2}}}\nonumber \\
 & +2e^{-\frac{s^{2}}{2}}Re[e^{2isIm(\alpha)}(1-\vert\langle A\rangle_{w}\vert^{2}-2iIm\langle A\rangle_{w}))\times\nonumber \\
 & (\alpha+\frac{s}{2})^{\ast}(\alpha-\frac{s}{2})]\}.\label{eq:21}
\end{align}
If we take $s=0$, then $\langle n\rangle_{s=0}=\vert\alpha\vert^{2}$
, this is the photon number for initial coherent state $\vert\alpha\rangle$.
We also can get the expectation value of $\langle a^{\dagger2}a^{2}\rangle$
as 

\begin{align}
\langle a^{\dagger2}a^{2}\rangle_{\Psi} & =\frac{\vert\lambda\vert^{2}}{4}\{\vert1+\langle A\rangle_{w}\vert^{2}\vert\alpha+\frac{s}{2}\vert^{4}+\vert1\!-\langle A\rangle_{w}\vert^{2}\vert\alpha-\frac{s}{2}\vert^{4}\nonumber \\
 & +2e^{-\frac{s^{2}}{2}}Re[e^{2is\vert\alpha\vert\sin\phi}(1+\langle A\rangle_{w}^{\ast})(1-\langle A\rangle_{w})\times\nonumber \\
 & (\alpha+\frac{s}{2})^{\ast2}(\alpha-\frac{s}{2})^{2}]\}\label{eq:22}
\end{align}
If we take $s=0$ , it gives the value under the initial state $\vert\alpha\rangle$,
i.g., $\langle a^{\dagger}a^{\dagger}aa\rangle_{s=0}=\vert\alpha\vert^{4}$.
By substituting the above expressions to the Eqs. (\ref{eq:13}) and
(\ref{eq:14}), we obtain the concrete expressions of $g_{coh}^{(2)}(0)$
and $Q_{m,coh}$, respectively. To investigating the effect of postselected
measurement on the photon statistical properties of coherent state
we plotted the analytical figures of $g_{coh}^{(2)}(0)$ and $Q_{m,coh}$,
and the results are shown in Fig. \ref{fig:2}. 

From Fig. \ref{fig:2} we can deduce that in postselected weak mesaurement
region with large weak values (see Fig. \ref{fig:fig1}(b) and (c)),
the final state of the pointer state possessed sub-Poisson statistics
$(0<g^{(2)}(0)<1$ and $-1<Q_{m}<0)$. From Fig. \ref{fig:fig1}(d),
we can see that in strong postselected measurement the system has
super Poisson statistics $(g^{(2)}(0)>1$ and $Q_{m}>0)$. Thus, the
results are summarized in Fig. \ref{fig:2} confirm that the postselected
measurement can change the statistical properties of the coherent
state significantly. 

\begin{figure}
\includegraphics[width=4cm]{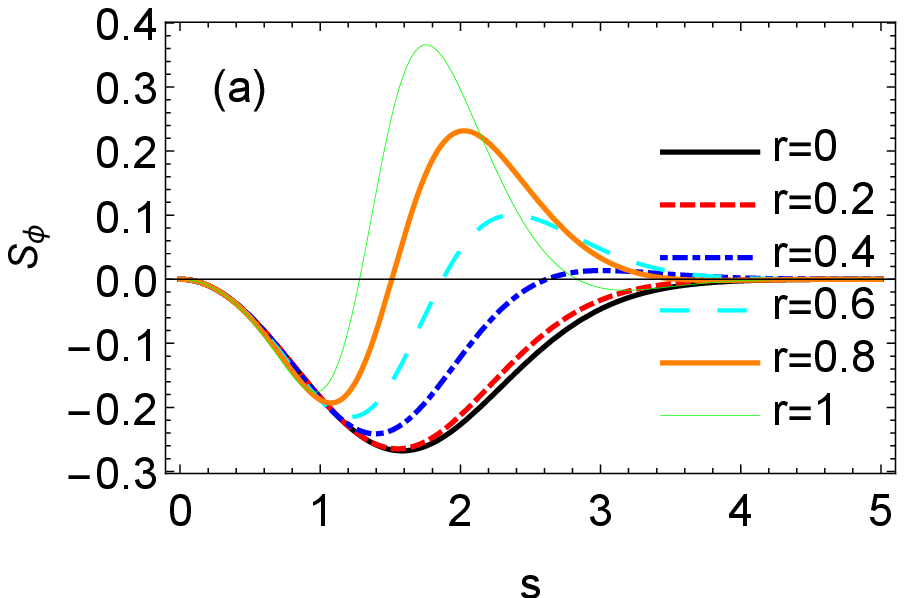}\includegraphics[width=4cm]{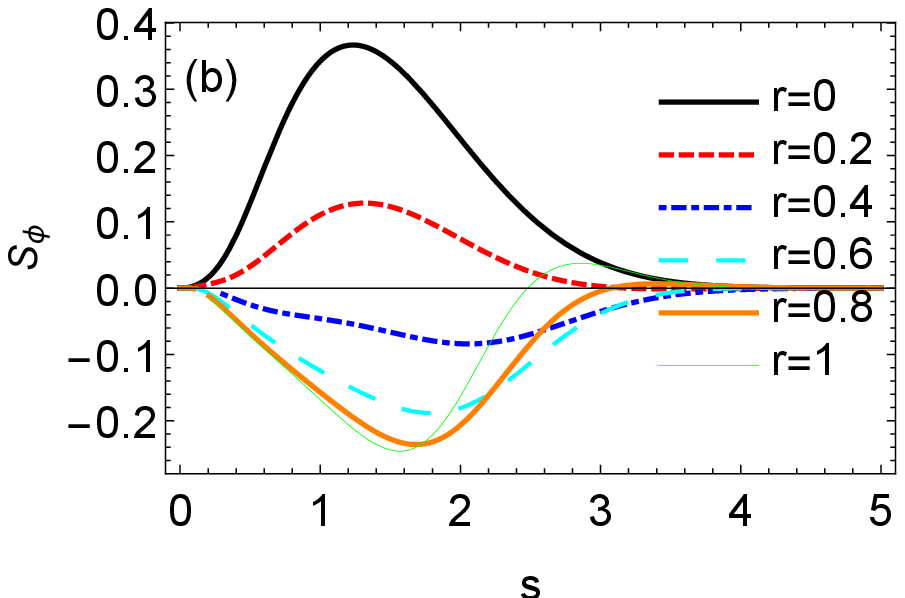}

\includegraphics[width=4cm]{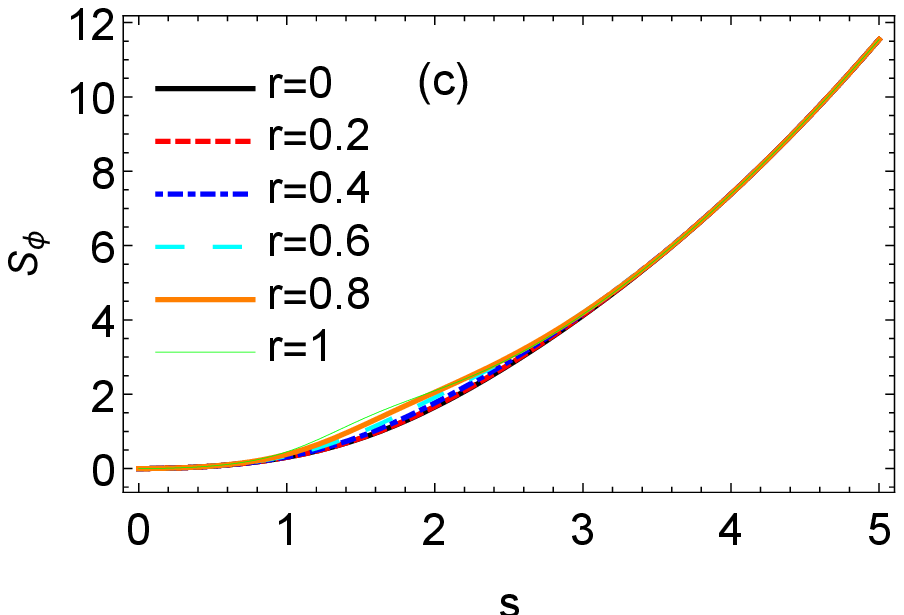}\includegraphics[width=4cm]{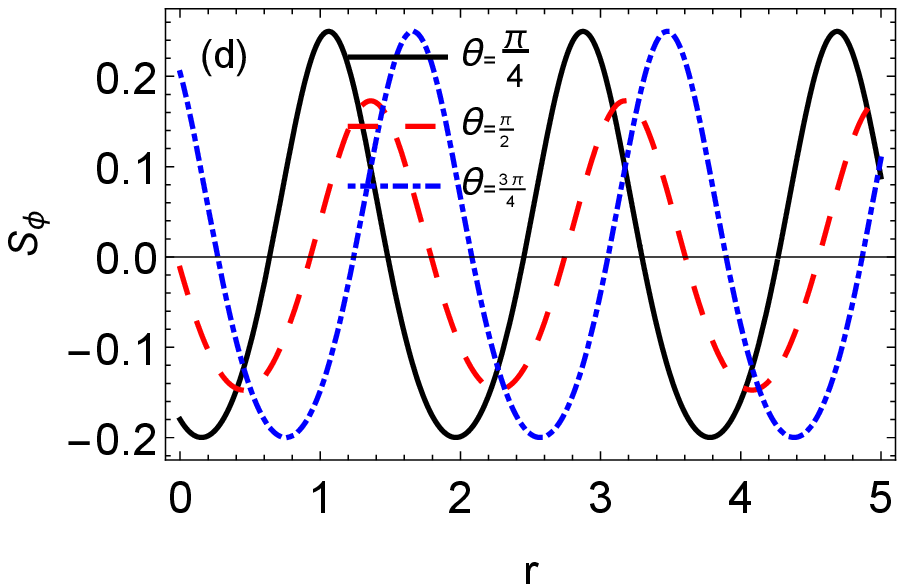}

\caption{\label{fig:3}(Color online) The Squeezing parameter $S_{\phi,coh}$
for coherent state after postselected measurement. Here, $\varphi=\frac{4\pi}{5}$,
$\vartheta=\frac{\pi}{3}$. $S_{\phi,coh}$vs interaction strength
$s$ for various $r$ and $P$ quadrature of coherent state ($\phi=\frac{\pi}{2}$),
but with different weak value: (a) $\theta=\frac{\pi}{9}$; (b) $\theta=\frac{7\pi}{9}$.
(c) $S_{\phi,coh}$ plotted as function of interaction strength $s$
for various $r$ and for $\theta=\frac{\pi}{9}$, $\phi=0$ (represents
the $X$ quadrature of coherent state). (d) $S_{\phi,coh}$ vs coherent
state parameter $r$ for different weak values and for $\phi=\frac{\pi}{2}$,
$s=2$.}
\end{figure}

Since the coherent state is a minimum uncertainty state, there is
no squeezing effect for coherent state $\vert\alpha\rangle$, i.e.,
$S_{\phi}=0$. Next, we will investigate the squeezing parameter $S_{\phi}^{coh}$
for $\vert\text{\ensuremath{\Psi}}\rangle$ of coherent pointer state.
The expectation value of $X_{\phi,coh}$ under the state $\vert\Psi\rangle$
is given by 
\begin{align}
\langle X_{\phi,coh}\rangle_{\Psi} & =\frac{\vert\lambda\vert^{2}}{\sqrt{2}}\{(1\!+\vert\langle A\rangle_{w}\vert^{2})\vert\alpha\vert\cos(\phi\!-\!\theta)\!+s\cos\phi Re[\langle A\rangle_{w}]\nonumber \\
 & +\frac{1}{2}e^{-\frac{1}{2}s^{2}}\!\!Re[e^{2isIm(\alpha)}(1-\langle A\rangle_{w})(1+\langle A\rangle_{w}^{\ast})\times\nonumber \\
 & (2r\cos(\vartheta-\phi)+is\sin\phi)].\label{eq:23-1}
\end{align}
If $s=0,$then the above expression reduced to the expectation value
of $X_{\phi}$ under the initial coherent pointer state $\vert\alpha\rangle$,
i.e., $\langle X_{\phi,coh}\rangle_{s=0}=Re[\alpha e^{-i\phi}]$.
The expectation value of $X_{\phi,coh}^{2}$ with $\vert\Psi\rangle$
is given by 
\begin{align}
\langle X_{\phi,coh}^{2}\rangle_{\Psi} & =\frac{1}{2}\langle(ae^{-i\phi}+a^{\dagger}e^{i\phi})(ae^{-i\phi}+a^{\dagger}e^{i\phi})\rangle\nonumber \\
 & =\frac{1}{2}\left[\langle a^{2}\rangle e^{-2i\phi}+\langle a^{\dagger2}\rangle e^{2i\phi}+2\langle a^{\dagger}a\rangle+1\right]\nonumber \\
 & =\frac{1}{2}\left[2Re[\langle a^{2}\rangle e^{-2i\phi}]+2\langle n\rangle+1\right]\label{eq:23}
\end{align}
where 
\begin{align}
\langle a^{2}\rangle & =\frac{\vert\lambda\vert^{2}}{4}\{\vert1+\langle A\rangle_{w}\vert^{2}(\alpha+\frac{s}{2})^{2}+\vert1-\langle A\rangle_{w}\vert^{2}(\alpha-\frac{s}{2})^{2}\nonumber \\
 & +e^{2isIm(\alpha)}e^{-\frac{s^{2}}{2}}(1-\langle A\rangle_{w})(1+\langle A\rangle_{w}^{\ast})(\alpha-\frac{s}{2})^{2}\nonumber \\
 & +e^{-2isIm(\alpha)}e^{-\frac{s^{2}}{2}}(1+\langle A\rangle_{w})(1-\langle A\rangle_{w}^{\ast})(\alpha+\frac{s}{2})^{2}\}.\label{eq:24-1}
\end{align}
 The squeezing parameter $S_{\phi,coh}$ of state $\vert\Psi\rangle$
can be obtained by substituting the Eq. (\ref{eq:23-1}) and Eq. (\ref{eq:23})
into Eq. (\ref{eq:17}), and its analytical results listed in Fig.
\ref{fig:3}. 

We can observe from Fig. \ref{fig:3} that after postselected measurement
the squeezing parameter of the initial coherent state changed dramatically,
and can see the phase-dependent squeezing effect; the $X$ quadrature
($\phi=0$) of final state not squeezed (see Fig. \ref{fig:3}(c)),
but $P$ quadrature ($\phi=\frac{\pi}{2}$) of the final state have
a squeezing effect for moderate interaction strengths (see Fig. \ref{fig:3}(a),(b)
and (d)) with any weak values. 

\subsection{Squeezed vacuum state}

Our second pointer state is squeezed vacuum state. Squeezed states
of the radiation field are generated by degenerate parametric down
conversion in an optical cavity \citep{PhysRevLett.57.2520}. The
squeezed state have important applications in many quantum information
processing tasks, including gravitational wave detection \citep{PhysRevD.80.042006,MckenzieSqueezed,Grote2016High},
quantum teleportation \citep{PhysRevLett.80.869,PhysRevA.60.937}.
Suppose that measuring device initially prepared in the squeezed vacuum
state \citep{Scullyl1997} which is defined by 
\begin{equation}
\vert\xi\rangle=S(\xi)\left|0\right\rangle \label{eq:Ss}
\end{equation}
 where $S(\xi)=\exp(\frac{1}{2}\xi^{\ast}a^{2}-\frac{1}{2}\xi a^{\dagger2})$,
and $\xi=\eta e^{i\delta}$ is an arbitrary complex number with modulus
$\eta$ and argument $\delta\in[0,2\pi]$. As indicated in Eq. (\ref{eq:Ss}),
the squeezed vacuum state is generated by the action on the vacuum
state $\vert0\rangle$ of the squeezing operator $S(\xi)$. After
the postselected measurement processes as outlined in the Section.II
, the normalized final pointer state can be written as 
\begin{equation}
\vert\varphi\rangle=\frac{\kappa}{2}[(1+\langle A\rangle_{w})\vert\xi,\frac{s}{2}\rangle+(1-\langle A\rangle_{w})\vert\xi,-\frac{s}{2}\rangle],
\end{equation}
where the normalization coefficient $\kappa$ is given by

\begin{equation}
\kappa=\sqrt{2}[1+\vert\langle A\rangle_{w}\vert^{2}+(1-\vert\langle A\rangle_{w}\vert^{2})e^{-\frac{1}{2}s^{2}\vert\cosh\eta+e^{i\delta}\sinh\eta\vert^{2}}]^{-\frac{1}{2}}
\end{equation}
and we note that $\vert\xi,\pm\frac{s}{2}\rangle=D(\pm\frac{s}{2})S\left(\xi\right)\vert0\rangle$
is squeezed coherent state. 

\begin{figure}
\includegraphics[width=4cm]{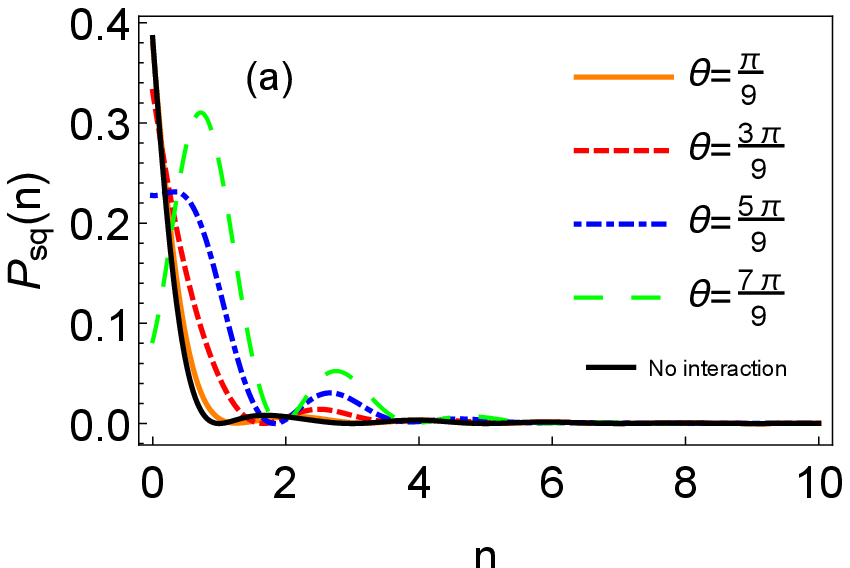}\includegraphics[width=4cm]{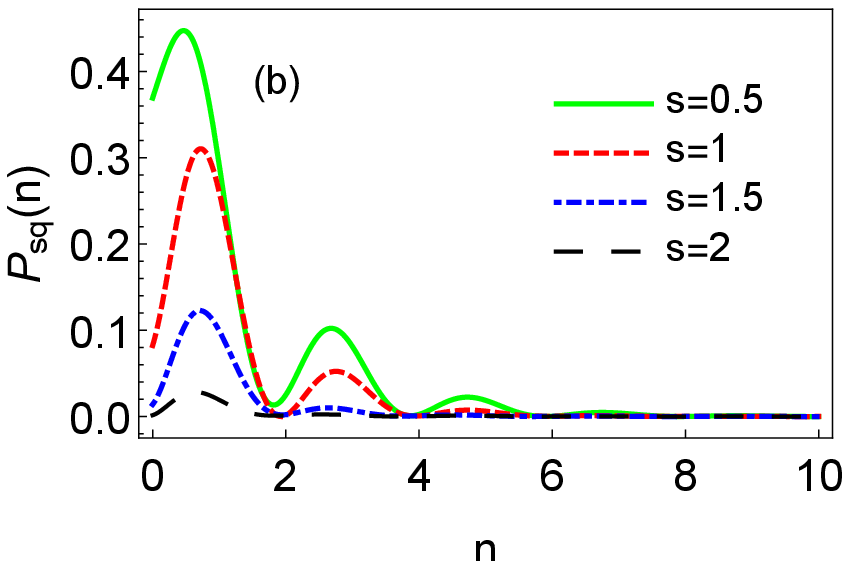}\caption{\label{fig:4}(Color online) Photon distribution $P_{sq}(n)$ of squeezed
vacuum state after postselected measurement as a function of $n$.
Here, $\delta=\frac{\pi}{3}$,$\varphi=\frac{\pi}{3}$, $\eta=0.5$.
(a) $P_{sq}(n)$ is plotted for $s=1$, and for no interaction case
(black curve) and various weak values. (b) $P_{sq}(n)$ is plotted
for $\theta=\frac{7\pi}{9}$, and for various interaction strength
$s$.}
\end{figure}

The probability amplitude of finding $n$ photons in a squeezed coherent
state is given by 
\begin{align}
\langle n\vert\pm\frac{s}{2},\xi\rangle & =\frac{1}{\sqrt{\cosh\eta}}\exp[-\frac{1}{2}\vert\pm\frac{s}{2}\vert^{2}-\frac{1}{2}(\pm\frac{s}{2})^{\ast2}e^{i\delta}\tanh\eta]\times\nonumber \\
 & \frac{\left(\frac{1}{2}e^{i\delta}\tanh\eta\right)^{\frac{n}{2}}}{\sqrt{n!}}H_{n}\left[\chi\left(e^{i\delta}\sinh(2r)\right)^{-\frac{1}{2}}\right]\label{eq:29}
\end{align}
with $\text{\ensuremath{\chi=\pm\frac{s}{2}\cosh\eta+(\pm\frac{s}{2})^{\ast}e^{i\delta}\sinh\eta}}$.
To find the conditional probability of $n$ photons under the state
$\vert\varphi\rangle$, we can use Eq. (\ref{eq:12}) by changing
$\vert\Phi\rangle$ to $\vert\varphi\rangle$, and use Eq. (\ref{eq:29}).
The effects of postselected measurement to the probability of finding
$n$ photons for state $\vert\varphi\rangle$ is displayed in Fig.
\ref{fig:4}. The black thick curve in Fig. \ref{fig:4}(a) corresponds
to the conditional photon probability for initial pointer state $\vert\xi\rangle$.
As illustrated in Fig. \ref{fig:4}, the postselected measurement
can change the photon distribution of the field, and in fewer photon
number region the $P_{sq}(n)$ with the state $\vert\varphi\rangle$
is larger than no interaction case (see Fig. \ref{fig:4}(a)). For
definite squeeze parameter $\eta$ and weak measurement region ($s<1$)
with large weak values, the $P_{qs}$is larger than strong measurement
region ($s>1$), but its occurred probability is small. 

The second-order correlation function $g^{(2)}(0)$ and Mandel factor
$Q_{m}$ of initial squeezed vacuum pointer state $\vert\xi\rangle$
is given by 
\begin{equation}
g^{(2)}(0)=3+\frac{1}{\sinh^{2}\eta}\label{eq:24}
\end{equation}
 and 

\begin{equation}
Q_{m}=1+2\sinh^{2}\eta,\label{eq:31-1}
\end{equation}
respectively. It is clear that the number fluctuations of squeezed
vacuum pointer state initially are super-Poissonian, and for all value
of $\eta$ both $g^{(2)}(0)$ and $Q_{m}$ can't take the values to
possess the sub-Poissonian statistics which is a nonclassical property
of the field. Now, we will study the effect of postselected measurement
to $g^{(2)}(0)$ and $Q_{m}$ of squeezed vacuum pointer state under
the normalized final pointer state $\vert\varphi\rangle$. The expectation
value of photon number operator $a^{\dagger}a$ and the $(a^{\dagger}a)^{2}$
under the final state $\vert\varphi\rangle$ can be calculated, and
their explicit expressions are given in Appendix. \ref{sec:App1}
(Since the expressions are too cumbersome to write here, we displayed
them in the Appendix ). 

\begin{figure}
\includegraphics[width=4cm]{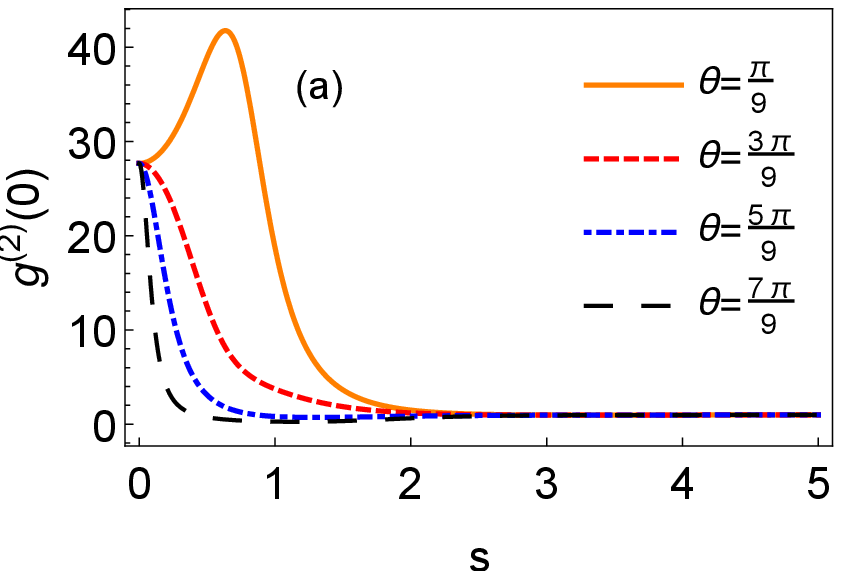}\includegraphics[width=4cm]{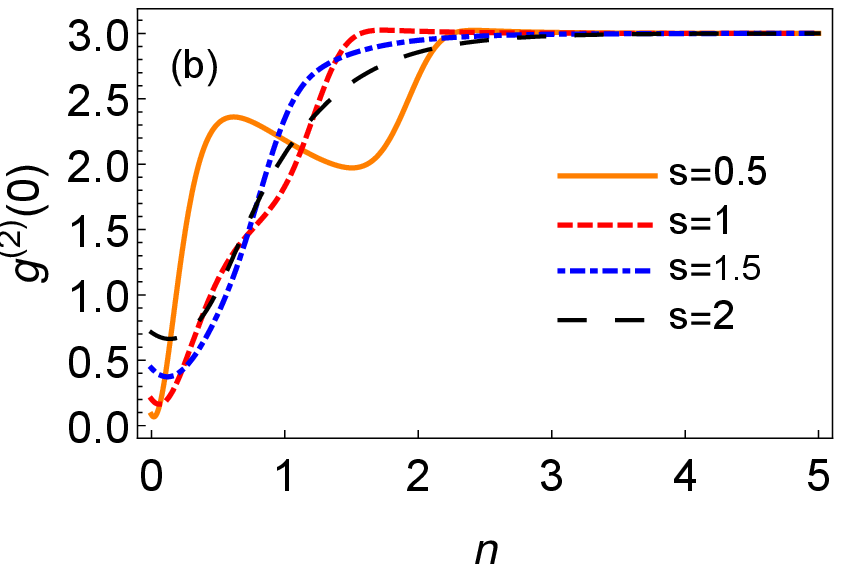}

\includegraphics[width=4cm]{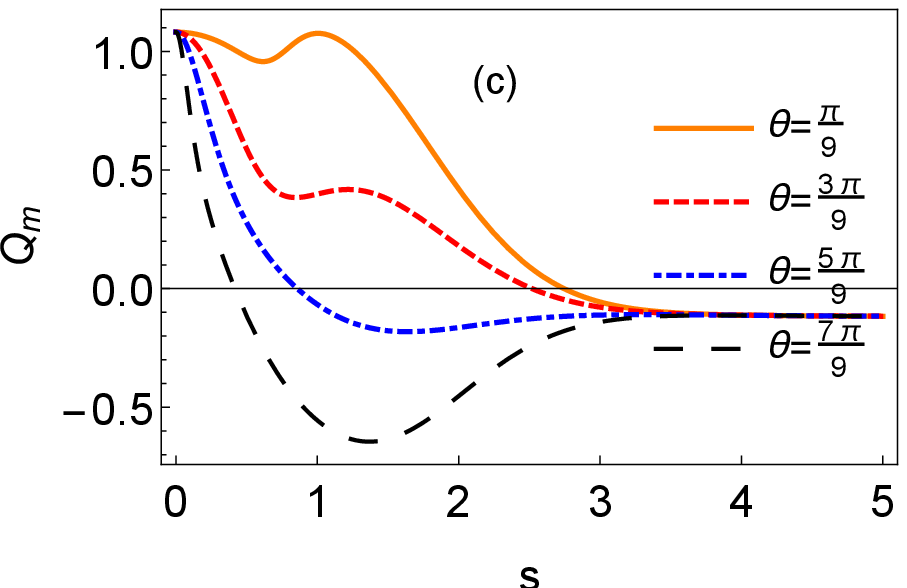}\includegraphics[width=4cm]{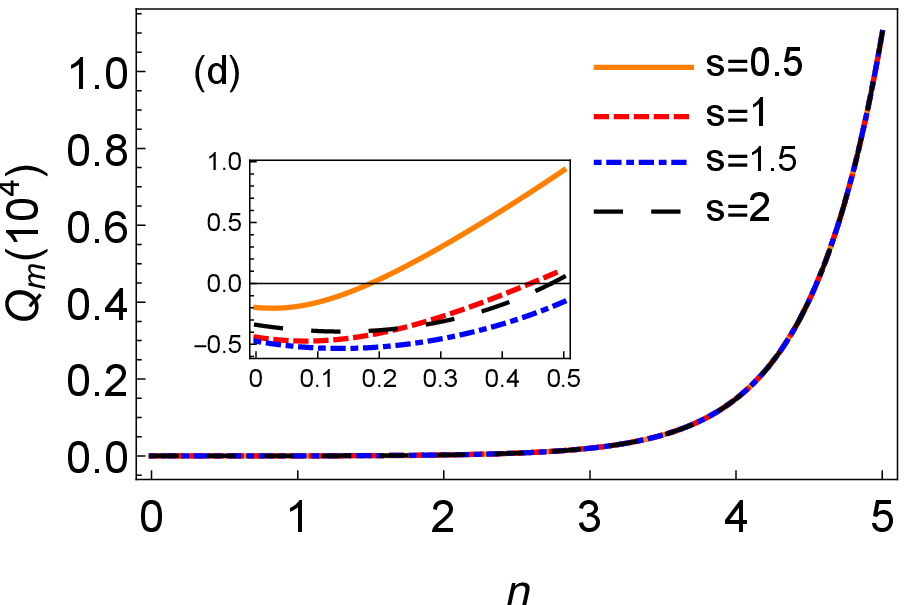}

\caption{\label{fig:5}(Color online) Second-order correlation function $g_{sq}^{(2)}(0)$
and Mandel factor $Q_{m,sq}$ of squeezed vacuum state after postselected
measurement. Here, $\varphi=\frac{\pi}{3}$,$\delta=\frac{\pi}{3}$.
(a) $g_{sq}^{(2)}(0)$ vs in interaction strength $s$ for different
weak values and for $\eta=0.2$. (b) $g_{sq}^{(2)}(0)$ plotted as
function of squeezed vacuum state parameter $\eta$ for various interaction
strength and for $\theta=\frac{7\pi}{9}$. (c) $Q_{m,sq}$ vs in interaction
strength $s$ for different weak values and for $\eta=0.2$. (d) $Q_{m,sq}$
plotted as function of squeezed vacuum state parameter $\eta$ for
various interaction strength and for $\theta=\frac{7\pi}{9}$, and
the inset in this figure depicts the curves in the interval $\eta\in[0,0.5]$.}
\end{figure}

The analytical results for $g_{sq}^{(2)}(0)$ and $Q_{m,sq}$ under
the state $\vert\varphi\rangle$ are presented in Fig. \ref{fig:5}.
It can be observed from Fig. \ref{fig:5}(a) and (c) that both $g_{sq}^{(2)}(0)$
and $Q_{m,sq}$ can take the values, which only sub-Poisson radiation
field can possess where the region $s\apprge0.5$ and for large weak
values. For definite measurement strength $s$ and larger weak value,
the value $g_{sq}^{(2)}(0)$ is lower than one and the value of $Q_{m,sq}$
is less than zero when the squeezed state parameter $\eta$ is less
than one (see Fig. \ref{fig:5}(b) and (d)). Thus, it is apparent
that the postselected measurement dramatically changed the photon
statistical properties of initial squeezed pointer state $\vert\xi\rangle$.

The squeezing parameter $S_{\phi}$ of initial squeezed vacuum state
$\vert\xi\rangle$ can be calculated by using Eqs. (\ref{eq:15}-\ref{eq:17})
and Eq. (\ref{eq:26}), and written as 

\begin{align}
S_{\phi} & =\frac{1}{2}\left[\cosh^{2}\eta-\sinh(2\eta)\cos(2\phi-\delta)+\sinh^{2}\eta\right]-\frac{1}{2}.\label{eq:31}
\end{align}
 It is evident from Eq. (\ref{eq:31}) that the squeezing effect of
squeezed vacuum state is phase-dependent: (i) if $\phi=\frac{\delta}{2}$,
then $S_{\phi}=-\frac{1}{2}(1-e^{-2\eta})$, it has squeezing effect
for $\eta>0$; (ii) if $\phi=\frac{\delta}{2}+\frac{\pi}{2}$ , then
$S_{\phi}=\frac{1}{2}(e^{2\eta}-1)$, there is no squeezing effect.
This reveals that it spread out the quadrature $X_{\phi=\frac{\delta}{2}}$
and at the same time squeezes the quadrature $X_{\phi=\frac{\delta}{2}+\frac{\pi}{2}}$.

\begin{figure}
\includegraphics[width=4cm]{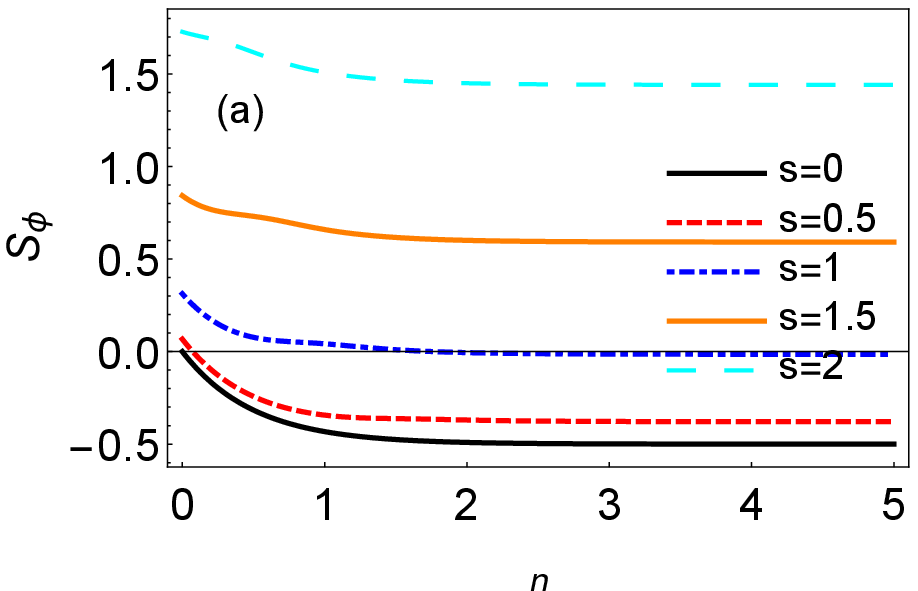}\includegraphics[width=4cm]{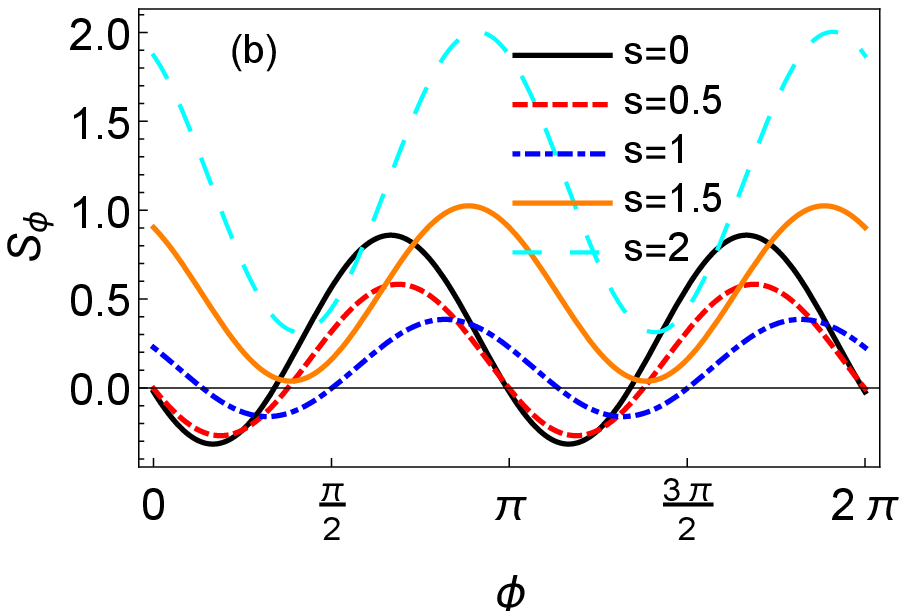}

\includegraphics[width=4cm]{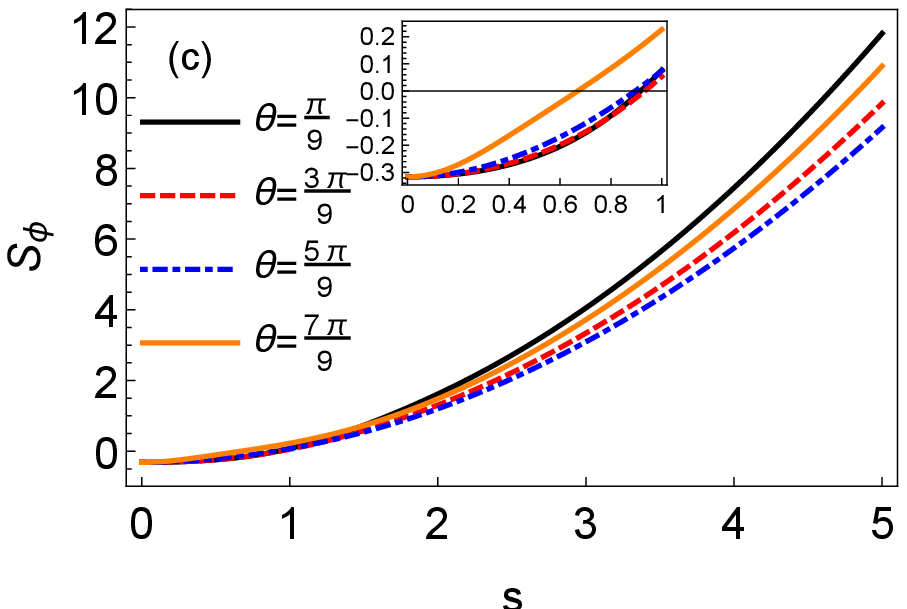}\includegraphics[width=4cm]{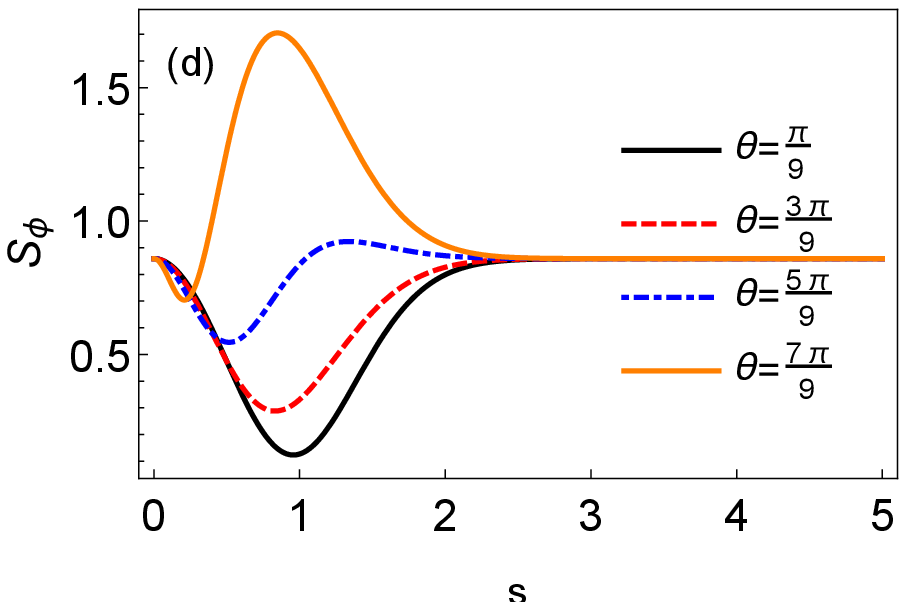}

\caption{\label{fig:6}(Color online) Squeezing parameter $S_{\phi,sq}$ of
squeezed vacuum state after postselected measurement. Here, $\varphi=\frac{\pi}{3}$.
(a) $S_{\phi}$ vs $\eta$ of squeezed vacuum state for different
interaction strength $s$ and for $\delta=\phi=0$, $\theta=\frac{\pi}{9}$.
(b)$S_{\phi,sq}$ plotted as a function of $\phi$ for various interaction
strength $s$ and for $\eta=0.5,\delta=\frac{\pi}{3}$, and $\theta=\frac{\pi}{9}$.
$S_{\phi,sq}$ vs interaction strength $s$ for different weak values
and for $\delta=0$ and $\eta=0.5$, but with different $\phi$: (c)
$\phi=0$; (d) $\text{\ensuremath{\phi=\frac{\pi}{2}}}$. The inset
in (c) depicts the curves in the interval $s\in(0,1]$.}
\end{figure}

Squeezing parameter $S_{\phi,sq}$ of final normalized state $\vert\varphi\rangle$
after postselected von-Nuemann measurement can be calculated as the
same processes of initial state $\vert\xi\rangle$, and the analytical
results showed in Fig. \ref{fig:6}. According to Fig. \ref{fig:6}
(a) and (b), the postselected measurement would have a negative effect
on the squeezing effect of squeezed vacuum state since $S_{\phi,sq}$
gradually becoming larger with increasing the measurement strength
$s$. However, as indicated in Fig. \ref{fig:6}(c), in weak measurement
regime where $0<s<1$ postselected measurement still has the effects
on the squeezing of squeezed vacuum pointer state. As mentioned earlier
the squeezing effect of squeezed vacuum state is phase-dependent,
and the postselected measurement has no positive effect on the squeezing
of $X_{\phi=\frac{\pi}{2}}$ quadrature (see Fig. \ref{fig:6}(d)). 

\subsection{Schrödinger cat state}

In the previous two subsections, we already confirmed that postselected
measurement characterized by a weak value can really change the statistical
and squeezing effect of single-mode radiation fields. To further increase
the reliability of our conclusions, in this subsection, we check the
same phenomena by taking the Schrödinger cat state as a pointer. Schrödinger\textquoteright s
cat is a gedanken experiment in quantum physics was proposed by Schrödinger
in 1935 \citep{E1935Die}, and its corresponding state is called Schrödinger
cat state. Schrödinger cat state play significant role not only in
fundamental tests of quantum theory \citep{PhysRevA.45.6811,PhysRevA.67.012105,PhysRevA.67.012106,PhysRevA.75.052105},
but also in many quantum information processing such as quantum computation
\citep{PhysRevA.68.042319}, quantum teleportation \citep{PhysRevA.64.022313,PhysRevA.64.052308}
and precision measurements \citep{PhysRevA.66.023819}. The Schrödinger
cat state is composed by the superposition of two coherent correlated
states moving in the opposite directions, and defined as \citep{Agarwal2013} 

\begin{equation}
\vert\varTheta\rangle=K\left(\vert\alpha\rangle+e^{i\omega}\vert-\alpha\rangle\right)\label{eq:26}
\end{equation}
 where
\begin{equation}
K=\left[2+2e^{-2\vert\alpha\vert^{2}}\cos\omega\right]^{-\frac{1}{2}}\label{eq:27}
\end{equation}
is normalization constant and $\text{\ensuremath{\alpha=re^{i\delta}}}$
is a coherent state parameter with modulus $r$ and argument $\delta$.
The normalized final state of Schrödinger cat state after the postselected
weak measurement is given by taking $\vert\phi\rangle=\vert\varTheta\rangle$
in Eq. (\ref{eq:6}), i.e., 
\begin{align}
\vert\chi\rangle & =\frac{\kappa}{2}\left[\left(1+\langle A\rangle_{w}\right)D\left(\frac{s}{2}\right)+\left(1-\langle A\rangle_{w}\right)D\left(-\frac{s}{2}\right)\right]\left|\Theta\right\rangle \label{eq:28}
\end{align}
 with normalization coefficient 
\begin{align}
\kappa^{-2} & =\frac{1}{2}(1+\vert\langle A\rangle_{w}\vert^{2})+K^{2}(1-\vert\langle A\rangle_{w}\vert^{2})\cos(2sIm[\alpha])e^{-\frac{s^{2}}{2}}\nonumber \\
 & +\frac{K^{2}}{2}\Re[(1-\langle A\rangle_{w})(1+\langle A\rangle_{w}^{\ast})\times\nonumber \\
 & (e^{i\omega}e^{-\frac{1}{2}\vert2\alpha+s\vert^{2}}+e^{-i\omega}e^{-\frac{1}{2}\vert2\alpha-s\vert^{2}})].
\end{align}

\begin{figure}
\includegraphics[width=4cm]{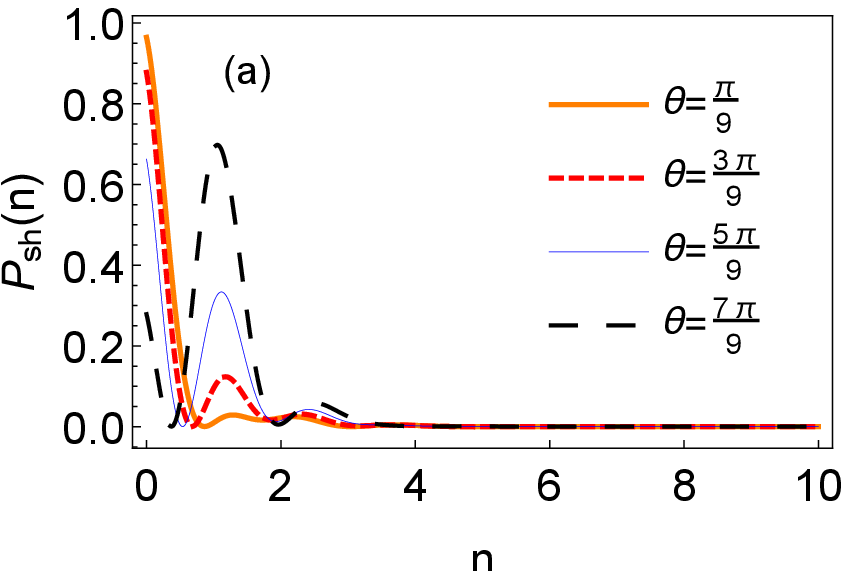}\includegraphics[width=4cm]{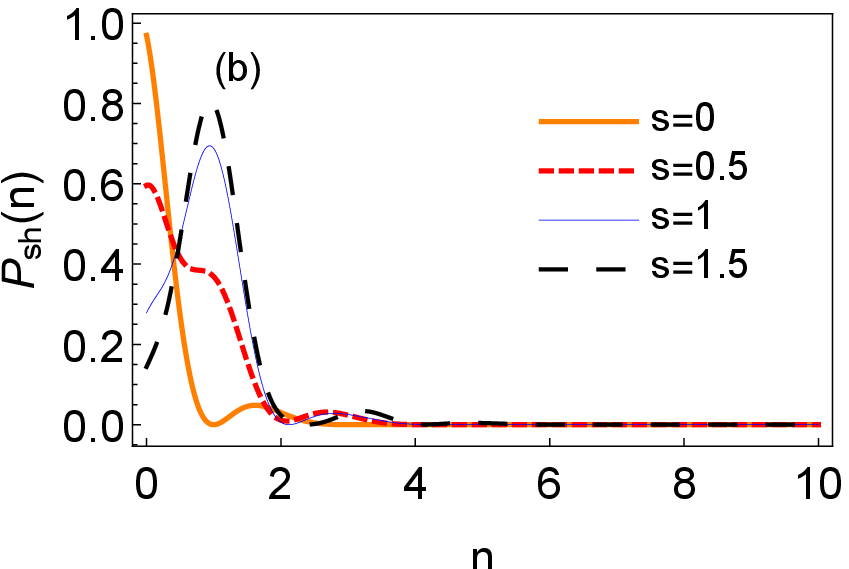}

\caption{\label{fig:7}(Color online) Photon distribution $P_{sh}(n)$ of squeezed
state as a function of photon number $n$. Here, $\delta=\frac{\pi}{3}$,$\varphi=\frac{\pi}{3}$,$\text{\ensuremath{\omega=0}}$,
$r=0.5$ (a) $P_{sh}(n)$ is plotted for $s=1$, and for various weak
values. (b) $P_{sh}(n)$ is plotted for $\theta=\frac{7\pi}{9}$,
and for various interaction strength $s$.}
\end{figure}

Here, we have to mention that $\omega\in[0,2\pi]$, and when $\omega=0$
($\text{\ensuremath{\omega=\pi}}$) it is called even (odd) Schrödinger
cat state. Similarly to the previous two cases, the conditional probability
$P_{sh}(n)$ of Schrödinger cat state can be obtained by replacing
the $\vert\phi\rangle$ in Eq. (\ref{eq:6}) to $\vert\chi\rangle$,
and the analytical results of even Schrödinger cat state is presented
in Fig. \ref{fig:7}. As shown in Fig. \ref{fig:7}, the postselected
measurement can change photon distribution of the field, and in fewer
photon regions with large weak values, $P_{sh}(n)$ larger than initial
photon distribution of Schrodinger even cat state. By comparing Fig.
\ref{fig:7} and Fig. \ref{fig:4} we also can get the same common
sense that the even Schrödinger cat state have similar properties
as a squeezed state. 

The Mandel factor and second-order correlation function for Schrodinger
cat state are given by 

\begin{align}
Q_{m} & =\frac{4\vert\alpha\vert^{2}e^{-2\vert\alpha\vert^{2}}\cos\omega}{1-e^{-4\vert\alpha\vert^{2}}\cos^{2}\omega}\label{eq:36}
\end{align}
and 
\begin{align}
g^{(2)}(0) & =1+\frac{4e^{-2\vert\alpha\vert^{2}}\cos\omega}{(1-e^{-2\vert\alpha\vert^{2}}\cos\omega)^{2}},\label{eq:37}
\end{align}
 respectively. It is apparent that both $g^{(2)}(0)$ and $Q_{m}$
have sub-Poisson statistics when $\cos\omega<0$ ($\frac{\pi}{2}<\omega<\frac{3\pi}{2}$). 

\begin{figure}
\includegraphics[width=4cm]{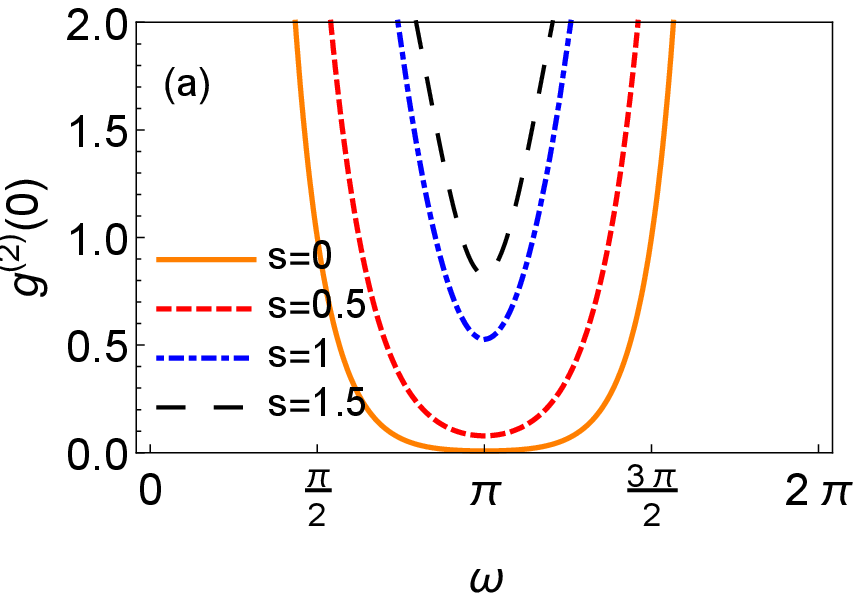}\includegraphics[width=4cm]{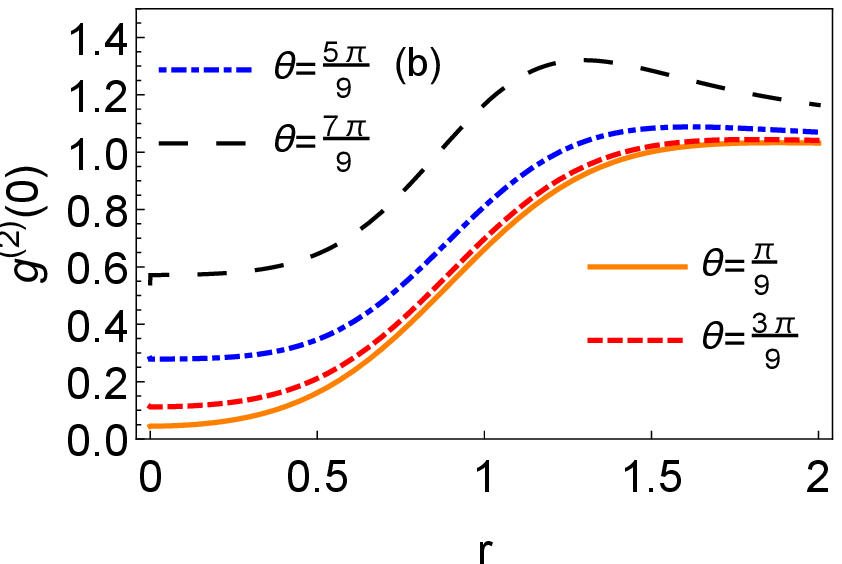}

\includegraphics[width=4cm]{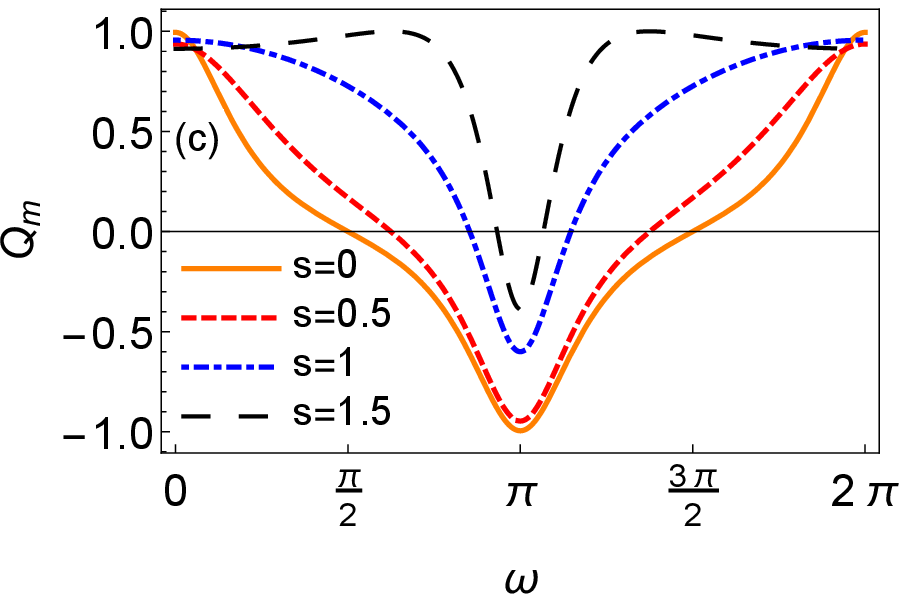}\includegraphics[width=4cm]{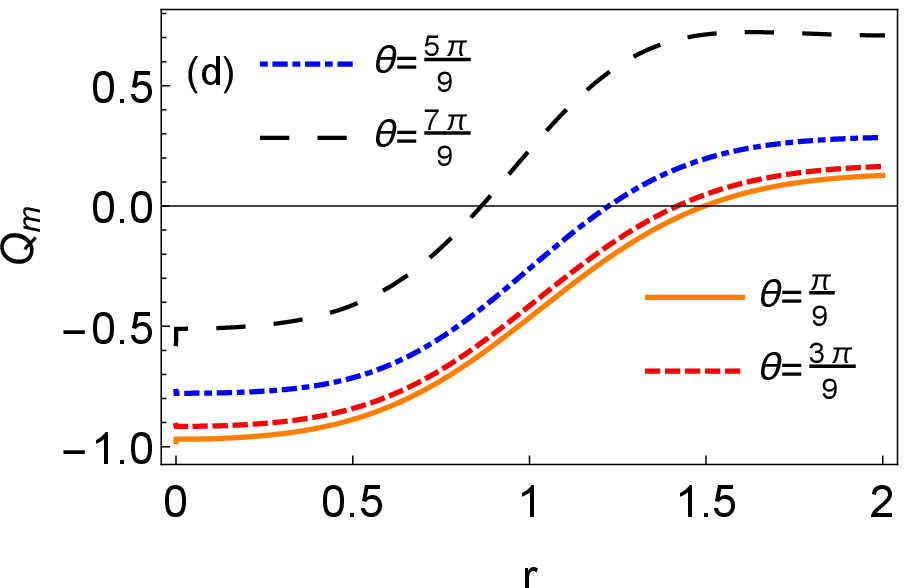}

\caption{\label{fig:8}(Color online) second-ordercorrelation function $g_{sh}^{(2)}(0)$
and Mandel factor $Q_{m,sh}$ of Schrödinger state after postselected
measurement. Here $\varphi=0$, $\delta=0$. (a) $g_{sh}^{(2)}(0)$
plotted as function $\omega$ of Schrödinger cat state for various
interaction strength $s$ and for $\theta=\frac{\pi}{9}$, $r=0.3$.
(b) $g_{sh}^{(2)}(0)$ vs parameter $r$ of Schrödinger cat state
for different weak values and for and $s=0.5$, $\omega=\pi$. (c)
$Q_{m,sh}$ plotted as function $\omega$ of Schrödinger cat state
for various interaction strength $s$ and for $\theta=\frac{\pi}{9}$,
$r=0.3$. (d) $Q_{m,sh}$ vs parameter $r$ of Schrödinger cat state
for different weak values and for $s=0.5$, $\omega=\pi$.}
\end{figure}

The $g_{sh}^{(2)}(0)$ and $Q_{m,sh}$ of Schrödinger cat state can
be calculated by using Eq. (\ref{eq:13}) and Eq. (\ref{eq:14}) with
normalized final pointer state $\vert\chi\rangle$. As indicated in
Fig. \ref{fig:8}, after postselected measurement, the statistical
property of Schrödinger cat state changed from sub-Poisson to super-Poisson
gradually with increasing the interaction strength and weak value.
However, as we can see in Fig. \ref{fig:8}(b) and (d), in postselected
weak measurement regime, where $0<s<1$, would give the best performance
for nonclassicality of odd Schrödinger cat state with small coherent
state modulus $r$. 

\begin{figure}
\includegraphics[width=4cm]{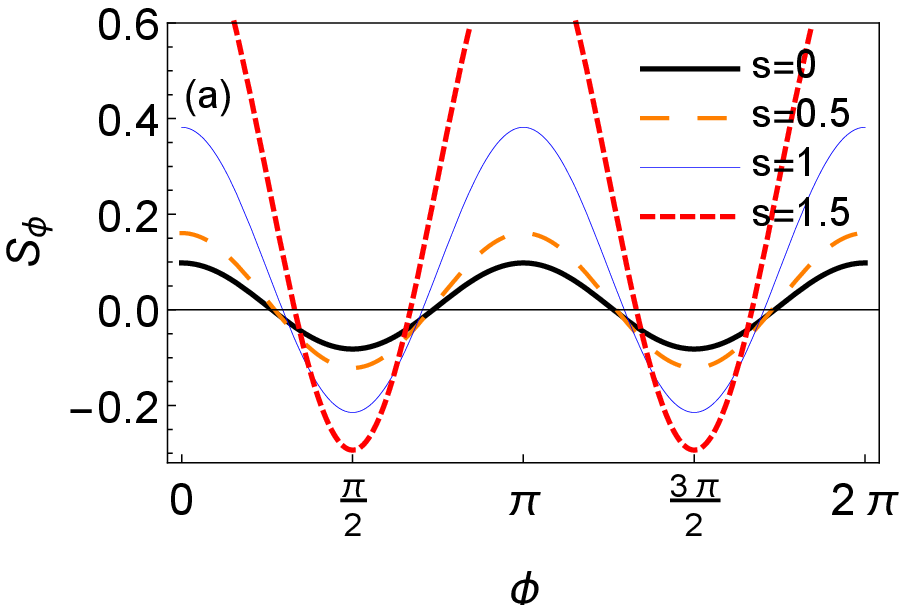}\includegraphics[width=4cm]{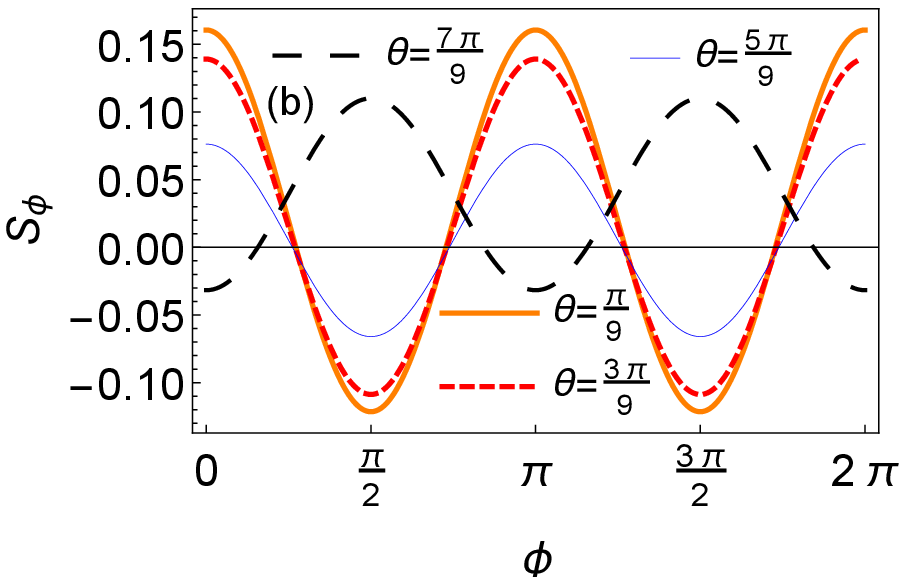}

\includegraphics[width=4cm]{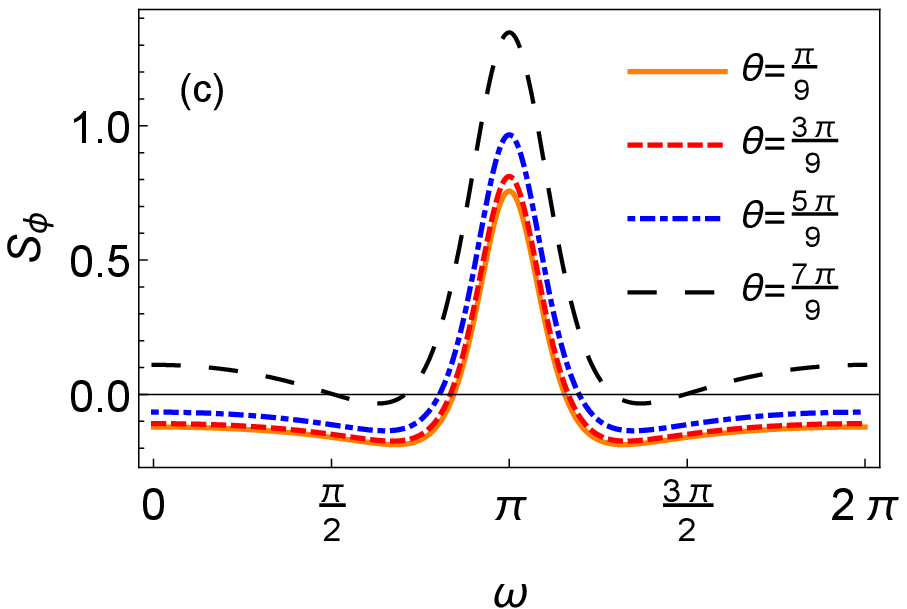}\includegraphics[width=4cm]{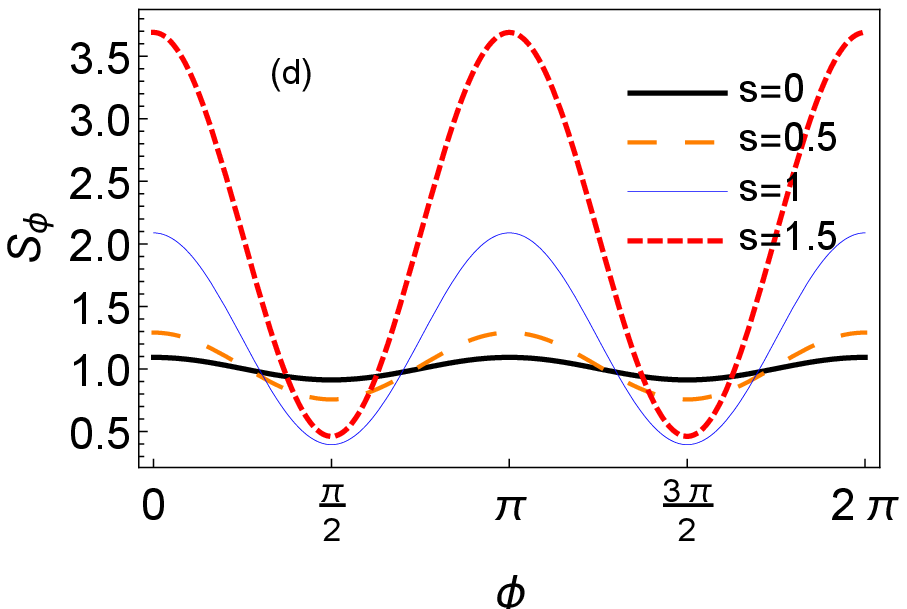}

\caption{\label{fig:9}(Color online) Squeezing parameter $S_{\phi,sh}$ of
Schrödinger cat state. Here $\varphi=0$, $\delta=0$, $r=0.3$. $S_{\phi,sh}$
of even Schrödinger cat state ($\omega=0$) vs squeezing parameter
angle $\phi$ for different interaction strength and for $\theta=\frac{\pi}{9}$
in (a); for different weak values and for $s=0.5$ in (b), respectively.
(c) $S_{\phi,sh}$ plotted as a function of $\omega$ for various
weak values and for $s=0.5$, $\phi=\frac{\pi}{2}$. (d) $S_{\phi,sh}$of
odd Schrödinger cat state ($\omega=\pi$) vs squeezing parameter angle
$\phi$ for different interaction strength and for $\theta=\frac{\pi}{9}.$}
\end{figure}

The squeezing parameter of initial Schrödinger cat state $\vert\varTheta\rangle$
reads as \citep{Agarwal2013}
\begin{align}
S_{\phi}^{in} & =\frac{\vert\alpha\vert^{2}e^{-4\vert\alpha\vert^{2}}}{(1+e^{-2\vert\alpha\vert^{2}}\cos\omega)^{2}}\times\nonumber \\
 & \left[1+\left(\cos2\phi(e^{2\vert\alpha\vert^{2}}+\cos\omega)^{2}+\sin^{2}\omega\cos^{2}\phi-1\right)\right]
\end{align}
 From this expression, we can deduce that the squeezing effect of
initial Schrödinger cat state is also phase-dependent similar to the
squeezed vacuum state, and the quadrature of the field will be squeezed
only when the inequality $\cos2\phi<\frac{\sin^{2}\omega\cos^{2}\phi-1}{(\cos\omega+e^{2\vert\alpha\vert^{2}})^{2}}$
satisfied. The details of this squeezing effect presented in Fig.
\ref{fig:9} (a)(see the black solid curve). 

The analytic expression of $S_{\phi,sch}$ for Schrödinger cat state
after postselected measurement can be achieved using final normalized
state $\vert\chi\rangle$, and the analytical results are summarized
in Fig. (\ref{fig:9}). As indicated in Fig. \ref{fig:9}, after postselected
measurement, the squeezing effect of Schrödinger cat state still depends
on phase $\phi$, and the squeezing effect of both even and odd Schrödinger
cat states are increased with increasing the interaction strength
$s$ and for small weak values (see Fig. \ref{fig:9} (a), (b) and
(d)). However, the odd Schrödinger cat pointer state has no squeezing
effect (see Fig. \ref{fig:9} (c) and (d)).

\section{\label{sec:IV}Conclusion and remarks }

In summary, we investigated the effects of postselected measurement
characterized by postselection and weak value on the statistical properties
and squeezing effects of single-mode radiation fields. To achieve
our goal, we take the coherent state, squeezed vacuum state and Schrödinger
cat state as a pointer, and their polarization degrees of freedom
as measures system, respectively. We derived analytical expressions
of those pointer state's normalized final state after postselected
measurement considering all interaction strengths between pointer
and measured system. We separately presented the exact expressions
of photon distributions, second-order correlation functions, Mandel
factors and squeezing parameters of the above three typical single-mode
radiation fields for corresponding final pointer states, and plotted
the figures to analyze the results. 

We found that the photon distributions of those three pointer states
changed significantly after postelected measurement, especially the
coherent pointer state. We showed that postselected measurement changed
the photon statistics and squeezing effect of coherent state dramatically,
and noticed that the amplification effect of weak value played a major
role in this process. We also showed that the postselected measurement
changed the photon statistics of squeezed vacuum state from super-Poisson
to sub-Poisson for large weak values, and moderate interaction strengths.
Whereas, the photon statistics of Schrödinger cat state changed from
sub-Poisson to super-Poisson with increasing the interaction strength.
In accordance with previous findings, the squeezing effects of squeezed
vacuum and Schrodinger cat pointer states are still, phase-dependent
and the squeezing effect of squeezed vacuum pointer is decreased with
increasing the interaction strength. On the contrary, the squeezing
effect of even Schrödinger cat state is increased with increasing
the interaction strength for small weak values compared with the initial
pointer state case. 

This work belong to the state optimization by using postselected von-neumann
measurement. Those properties of radiation fields we investigated
in this paper, are directly effect on the implementations of the important
quantum information processing which mentioned in introduction part.
Thus, we anticipate that the results showed in this research can be
useful to provide other effective methods for studying the related
quantum information processing. 

In our current research, we only consider the three typical single-mode
radiation fields to investigate the effects of postselected von-Neuman
measurement on their inherent properties, but real light beams are
in fact the time-dependent and the representation of time dependence
requires the use of two or more modes of the optical systems. Thus,
it would be interesting to check the effects of postselected von-neumann
measurement on other useful radiation fields in quantum optics and
quantum information processing such as single photon-added \citep{Zavatta660,PhysRevA.82.063833}
and -subtracted \citep{PhysRevA.75.032104} states, pair-coherent
state (two-mode field) \citep{agarwal1988nonclassical,PhysRevA.50.2865,Xin_1996,article}
and other multimode radiation fields \citep{article1}, respectively. 
\begin{acknowledgments}
This work was supported by the National Natural Science Foundation
of China (Grant No. 11865017, No.11864042).
\end{acknowledgments}

\appendix

\section{\label{sec:App1}The explicit expressions of some quantities}

\begin{widetext}

1. For squeezed vacuum pointer state: The expectation value of photon
number operator $a^{\dagger}a$ under the state $\vert\chi\rangle$
is given by 

\begin{equation}
\langle a^{\dagger}a\rangle_{\varphi}=\frac{\text{\ensuremath{\vert\kappa\vert^{2}}}}{4}\{2(1+\vert\langle A\rangle_{w}\vert^{2})(\frac{s^{2}}{4}+\sinh^{2}\eta)+2Re[(1-\text{\ensuremath{\langle A\rangle_{w}}})(1+\langle A\rangle_{w})^{\ast}I]\}
\end{equation}
with 
\[
I=\exp[-\frac{s^{2}}{2}\vert\cosh\eta+e^{i\delta}\sinh\eta\vert^{2}](\sinh^{2}\eta+\frac{s^{2}}{4}-\frac{s^{2}}{2}(1+i\sin\delta\sinh2\eta))
\]
 In calculating the squeezing parameter, we use $\text{\ensuremath{\langle X_{\phi}\rangle_{\varphi}}}$
and $\langle X_{\phi}^{2}\rangle_{\varphi}$, and their expressions
are given by 
\begin{align*}
\langle X_{\phi}\rangle_{\varphi} & =s\frac{\vert\kappa\vert^{2}}{4\sqrt{2}}\{\cos\phi\vert1+\langle A\rangle_{w}\vert^{2}-\cos\phi\vert1+\langle A\rangle_{w}\vert^{2}\\
 & +2\exp[-\frac{s^{2}}{2}\vert\cosh\eta+e^{i\delta}\sinh\eta\vert^{2}]Re[e^{-i\theta}(1+\langle A\rangle_{w})(1-\langle A\rangle_{w})^{\ast}(\cosh^{2}\eta+\frac{1}{2}e^{i\delta}\sinh2\eta-\frac{1}{2})]\\
 & -2\exp[-\frac{s^{2}}{2}\vert\cosh\eta+e^{i\delta}\sinh\eta\vert^{2}]Re\{e^{i\theta}(1+\langle A\rangle_{w})^{\ast}(1-\langle A\rangle_{w})[\cosh^{2}\eta+e^{i\delta}\frac{1}{2}\sinh2\eta-\frac{1}{2}]\}
\end{align*}
 and 

\begin{align}
\langle X_{\theta}^{2}\rangle & =\frac{1}{2}[2Re(IIe^{-2i\phi})+2\langle a^{\dagger}a\rangle+1]
\end{align}
 with 
\begin{align*}
II & =\frac{\vert\kappa\vert^{2}}{4}\{\left(\frac{s^{2}}{4}-\frac{1}{2}\sinh(2\eta)e^{-i\delta}\right)\vert1+\langle A\rangle_{w}\vert^{2}+\left(\frac{s^{2}}{4}-\frac{1}{2}\sinh(2\eta)e^{-i\delta}\right)\vert1-\langle A\rangle_{w}\vert^{2}\\
 & +(1-\vert\langle A\rangle_{w}\vert^{2})III\}
\end{align*}
where 
\begin{align*}
I & II=\frac{s^{2}}{4}+s^{2}(\frac{1}{2}e^{-i\delta}\sinh2\eta+\sinh^{2}\eta)+s^{2}(\cosh\eta+e^{i\delta}\sinh\eta)^{2}e^{-2i\delta}\sinh^{2}\eta\\
 & -e^{-i\delta}\frac{1}{2}\sinh2\eta\}\exp[-\frac{s^{2}}{2}\vert\cosh\eta+e^{i\delta}\sinh\eta\vert^{2}]
\end{align*}
 respectively. 

2. For Schrödinger cat pointer state: The expectation value of photon
number operator $a^{\dagger}a$ and $a^{\dagger2}a^{2}$ under the
state $\vert\chi\rangle$ is given by 
\begin{align}
\langle a^{\dagger}a\rangle_{\chi} & =\frac{\vert\kappa\vert^{2}K^{2}}{4}\{\vert1+\langle A\rangle_{w}\vert^{2}\left[\vert\alpha+\frac{s}{2}\vert^{2}+\vert-\alpha+\frac{s}{2}\vert^{2}+e^{i\omega}(\alpha+\frac{s}{2})^{\ast}(-\alpha+\frac{s}{2})e^{-2\vert\alpha\vert^{2}}+e^{-i\omega}(-\alpha+\frac{s}{2})^{\ast}(\alpha+\frac{s}{2})e^{-2\vert\alpha\vert^{2}}\right]\nonumber \\
 & +\vert1-\langle A\rangle_{w}\vert^{2}\left[\vert\alpha-\frac{s}{2}\vert^{2}+\vert\alpha+\frac{s}{2}\vert^{2}-e^{i\omega}(\alpha-\frac{s}{2})^{\ast}(\alpha+\frac{s}{2})e^{-2\vert\alpha\vert^{2}}-e^{-i\omega}(\alpha+\frac{s}{2})^{\ast}(\alpha-\frac{s}{2})e^{-2\vert\alpha\vert^{2}}\right]\nonumber \\
 & -2Re[(e^{i\omega}\vert\alpha+\frac{s}{2}\vert^{2}e^{-2\vert\alpha+\frac{s}{2}\vert^{2}}+e^{-i\omega}e^{-2\vert\alpha-\frac{s}{2}\vert^{2}}\vert\alpha-\frac{s}{2}\vert^{2}-2e^{-\frac{s^{2}}{2}}Re[e^{2isIm(\alpha)}(\alpha+\frac{s}{2})^{\ast}(\alpha-\frac{s}{2})])\times\nonumber \\
 & (1-\langle A\rangle_{w})(1+\langle A\rangle_{w})^{\ast}]\},
\end{align}
 and 
\begin{align}
\langle a^{\dagger2}a^{2}\rangle_{\chi} & =\frac{\vert\kappa\vert^{2}K^{2}}{4}\{\vert1+\langle A\rangle_{w}\vert^{2}\left[\vert\alpha+\frac{s}{2}\vert^{4}+\vert-\alpha+\frac{s}{2}\vert^{4}+e^{i\omega}(\alpha^{\ast}+\frac{s}{2})^{2}(-\alpha+\frac{s}{2})^{2}e^{-2\vert\alpha\vert^{2}}+e^{-i\omega}(-\alpha^{\ast}+\frac{s}{2})^{2}(\alpha+\frac{s}{2})^{2}e^{-2\vert\alpha\vert^{2}}\right]\nonumber \\
 & +\vert1-\langle A\rangle_{w}\vert^{2}\left[\vert\alpha-\frac{s}{2}\vert^{4}+\vert\alpha+\frac{s}{2}\vert^{4}+e^{i\omega}e^{-2\vert\alpha\vert^{2}}(\alpha^{\ast}-\frac{s}{2})^{2}(\alpha+\frac{s}{2})^{2}+e^{-i\omega}e^{-2\vert\alpha\vert^{2}}(\alpha^{\ast}+\frac{s}{2})(\alpha-\frac{s}{2})^{2}\right]\nonumber \\
 & +2Re[(e^{i\omega}\vert\alpha+\frac{s}{2}\vert^{4}e^{-2\vert\alpha+\frac{s}{2}\vert^{2}}+e^{-i\omega}\vert\alpha-\frac{s}{2}\vert^{4}e^{-2\vert\alpha-\frac{s}{2}\vert^{2}}+2e^{-\frac{s^{2}}{2}}Re[e^{2isIm(\alpha)}(\alpha^{\ast}+\frac{s}{2})^{2}(\alpha-\frac{s}{2})^{2}])\times\nonumber \\
 & (1-\langle A\rangle_{w})(1+\langle A\rangle_{w})^{\ast}]\},
\end{align}
 respectively. In calculating the squeezing parameter, we use $\text{\ensuremath{\langle X_{\phi}\rangle_{\chi}}}$
and $\langle X_{\phi}^{2}\rangle_{\chi}$, and their expressions are
given by 
\begin{align*}
\langle X_{\phi}\rangle_{\chi} & =\sqrt{2}Re[\langle a\rangle_{\chi}e^{-i\phi}]
\end{align*}
 with 
\begin{align*}
\langle a\rangle_{\chi} & =\frac{\vert\kappa\vert^{2}K^{2}}{4}\{\vert1+\langle A\rangle_{w}\vert^{2}[s+e^{i\phi}(-\alpha+\frac{s}{2})e^{-2\vert\alpha\vert^{2}}+e^{-i\phi}(\alpha+\frac{s}{2})e^{-2\vert\alpha\vert^{2}}]\\
 & +\vert1-\langle A\rangle_{w}\vert^{2}[-s-e^{i\phi}(\alpha+\frac{s}{2})e^{-2\vert\alpha\vert^{2}}+e^{-i\phi}(\alpha-\frac{s}{2})e^{-2\vert\alpha\vert^{2}}]\\
 & +(1-\langle A\rangle_{w})(1+\langle A\rangle_{w})^{\ast}[(2i\alpha\sin(2sIm[\alpha])-s\cos(2sIm[\alpha]))e^{-\frac{1}{2}s^{2}}-e^{i\phi}(\alpha+\frac{s}{2})e^{-\frac{1}{2}\vert2\alpha+s\vert^{2}}+e^{-i\phi}(\alpha-\frac{s}{2})e^{-\frac{1}{2}\vert2\alpha-s\vert^{2}}]\\
 & +(1-\langle A\rangle_{w})^{\ast}(1+\langle A\rangle_{w})[(-2i\alpha\sin(2sIm[\alpha])+s\cos(2sIm[\alpha]))e^{-\frac{1}{2}s^{2}}+e^{i\phi}(-\alpha+\frac{s}{2})e^{-\frac{1}{2}\vert2\alpha-s\vert^{2}}+e^{-i\phi}(\alpha+\frac{s}{2})e^{-\frac{1}{2}\vert2\alpha+s\vert^{2}}]\}
\end{align*}
 and 
\begin{align*}
\langle X_{\phi}^{2}\rangle_{\chi} & =\frac{1}{2}[2\langle a^{\dagger}a\rangle_{\chi}+2Re[\langle a^{2}\rangle_{\chi}e^{-2i\phi}]+1]
\end{align*}
with 
\begin{align}
\langle a^{2}\rangle_{\chi} & =\frac{\vert\kappa\vert^{2}K^{2}}{4}\{\vert1+\langle A\rangle_{w}\vert^{2}\left[2(\alpha^{2}+\frac{s^{2}}{4})+e^{i\omega}e^{-2\vert\alpha\vert^{2}}(-\alpha+\frac{s}{2})^{2}+e^{-i\omega}e^{-2\vert\alpha\vert^{2}}(\alpha+\frac{s}{2})^{2}\right]\nonumber \\
 & +\vert1-\langle A\rangle_{w}\vert^{2}\left[2(\alpha^{2}+\frac{s^{2}}{4})+e^{i\omega}e^{-2\vert\alpha\vert^{2}}(\alpha+\frac{s}{2})^{2}+e^{-i\omega}e^{-2\vert\alpha\vert^{2}}(\alpha-\frac{s}{2})^{2}\right]\nonumber \\
 & \!\!+(1-\!\langle A\rangle_{w})(1\!+\!\langle A\rangle_{w})^{\ast}[2e^{-\frac{s^{2}}{4}}(\cos(2sIm[\alpha])(\alpha^{2}+\frac{s^{2}}{4})-\!is\alpha\sin(2sIm[\alpha]))\!+\!e^{-i\omega}(\alpha-\frac{s}{2})^{2}e^{-2\vert\alpha-\frac{s}{2}\vert^{2}}\!+e^{i\omega}(\alpha+\frac{s}{2})^{2}e^{-2\vert\alpha+\frac{s}{2}\vert^{2}}]\nonumber \\
 & \!\!+\!(1+\!\langle A\rangle_{w})(1-\!\langle A\rangle_{w})^{\ast}[2e^{-\frac{s^{2}}{4}}(\cos(2sIm[\alpha])(\alpha^{2}+\frac{s^{2}}{4})\!-\!is\alpha\sin(2sIm[\alpha]))\!+\!e^{i\omega}(\alpha-\frac{s}{2})^{2}e^{-2\vert\alpha-\frac{s}{2}\vert^{2}}\!+e^{-i\omega}(\alpha+\frac{s}{2})^{2}e^{-2\vert\alpha+\frac{s}{2}\vert^{2}}]\},
\end{align}
 respectively. 

\end{widetext}

\bibliographystyle{apsrev4-1}
\bibliography{ref}

\end{document}